\documentclass[aps,prd,preprint,tightenlines,superscriptaddress]{revtex4}
\usepackage{graphicx}
\usepackage{epstopdf}
\usepackage{subfigure}
\usepackage{psfrag}
\usepackage{amsmath,amssymb}
\usepackage{colordvi}
\usepackage{amsfonts}
\usepackage{enumerate}
\usepackage{slashed}
\usepackage{color}
\usepackage{xcolor}
\graphicspath{{./}{pandaxCDR_pmt/}}

\begin{document}

\title{PandaX: A Liquid Xenon Dark Matter Experiment at CJPL }

\date{\today}
\vspace{0.5in}
\newcommand{\sjtuphys}{\affiliation{INPAC and Department of Physics and Astronomy, \\Shanghai Jiao Tong University, Shanghai, 200240, P. R. China}}
\newcommand{\sjtume}{\affiliation{School of Mechanical Engineering, \\Shanghai Jiao Tong University, Shanghai, 200240, P. R. China}}
\newcommand{\sdu}{\affiliation{School of Physics and Key Laboratory of Particle Physics and Particle Irradiation (MOE), Shandong University, Jinan 250100, China}}
\newcommand{\sinap}{\affiliation{Shanghai Institute of Applied Physics, Shanghai, 201800, P. R. China}}
\newcommand{\umich}{\affiliation{Department of Physics, University of Michigan, Ann Arbor, MI, 48109, USA}}
\newcommand{\pku}{\affiliation{School of Physics, Peking University, Beijing, 100080, P. R. China}}
\newcommand{\umd}{\affiliation{Department of Physics, University of Maryland, College Park, MD, 20742, USA}}
\newcommand{\yalong}{\affiliation{Yalong River Hydropower Development Company, Ltd., 288 Shuanglin Road, Chengdu, 610051, P. R. China}}

\sjtuphys
\sjtume
\sdu
\sinap
\umich
\pku
\umd
\yalong

\author{X.~G.~Cao} \sinap
\author{X.~Chen} \sjtuphys
\author{Y.~H.~Chen} \yalong
\author{X.~Y.~Cui} \sjtuphys
\author{D.~Q.~Fang} \sinap
\author{C.~B.~Fu} \sjtuphys
\author{K.~L.~Giboni} \sjtuphys
\author{H.~W.~Gong} \sjtuphys
\author{G.~D.~Guo} \sjtuphys
\author{M.~He} \sjtuphys
\author{J.~Hu} \sjtuphys
\author{X.~T.~Huang} \sdu
\author{X.~D.~Ji} \thanks{Spokesperson, contact: xdji@sjtu.edu.cn and xji@umd.edu} \sjtuphys\pku\umd
\author{Y.~L.~Ju} \sjtume
\author{S.~L.~Li} \sjtuphys
\author{Q.~Lin} \sjtuphys
\author{H. X. Liu} \sjtume
\author{J.~L.~Liu} \sjtuphys
\author{X.~Liu} \sjtuphys
\author{W.~Lorenzon} \umich
\author{Y.~G.~Ma} \sinap
\author{Y.~J.~Mao} \pku
\author{K.~X.~Ni} \sjtuphys
\author{K.~Pushkin} \sjtuphys\umich
\author{X.~X.~Ren} \sdu
\author{M.~Schubnell} \umich
\author{M.~B.~Shen} \yalong
\author{Y.~J.~Shi} \sjtuphys
\author{S.~Stephenson} \umich
\author{A.~D.~Tan} \umd
\author{G.~Tarle} \umich
\author{H.~W.~Wang} \sinap
\author{J.~M.~Wang} \yalong
\author{M.~Wang} \sdu
\author{X.~M.~Wang} \sjtuphys
\author{Z.~Wang} \sjtume
\author{Y.~H.~Wei} \sjtuphys
\author{S.~Y.~Wu} \yalong
\author{M.~J.~Xiao} \sjtuphys
\author{X.~Xiao} \sjtuphys
\author{P.~W.~Xie} \sjtuphys
\author{T. Ye} \sjtuphys
\author{Y.~H.~You} \yalong
\author{X.~H.~Zeng} \yalong
\author{H.~Zhang} \sjtume
\author{T.~Zhang} \sjtuphys
\author{H. Y. Zhao} \sjtuphys
\author{L.~Zhao} \sjtuphys
\author{X. P. Zhou} \pku
\author{Z.~H.~Zhu} \yalong
\collaboration{PandaX Collaboration}

\date{\today}
\vspace{0.5in}

\begin{abstract}

PandaX is a large upgradable liquid-xenon detector system that can be used for both direct dark-matter detection and $^{136}$Xe double-beta decay search. It is located in the Jinping Deep-Underground Laboratory
in Sichuan, China. The detector operates in dual-phase mode, allowing detection of both prompt scintillation, and ionization charge through proportional scintillation. The central time projection chamber will be staged, with the first stage accommodating a target mass of about 120\,kg. In stage II, the target mass will be increased to about 0.5\,ton. In the final stage, the detector can be upgraded to a multi-ton target mass. In this paper a detailed description of the stage-I detector design and performance results established during the commissioning phase is presented.
\end{abstract}

\maketitle

\section{Introduction}
\label{sec:intro}

Overwhelming evidence from astrophysical observations indicates that only 20\%
of the matter in the universe is made of ordinary matter, while the remaining
80\% is made of some form of matter called ``dark matter''~\cite{Bertone:2004pz}. Although these
indirect astrophysical observations convince us that dark matter exists, dark
matter has not yet been directly observed. The standard model of particle
physics, which has been very successful in explaining the properties of
ordinary matter, can neither explain dark matter's existence nor its
properties.  The discovery and identification of dark matter would have a
profound impact on cosmology, astronomy, and particle physics. A leading dark
matter candidate consistent with all astrophysical data is a weakly interacting
massive particle (WIMP)~\cite{Goodman:1985,Jungman:1996,Freedman:2003}.
WIMPs could be studied in standard particle physics through either observations of ordinary matter particles produced through DM annihilations in the halo of the Milky Way,  production of DM particles through high-energy collisions in the accelerators such as the Large Hadron Collider (LHC), or WIMPs could be detected through their interactions with atomic nuclei in specially designed detectors. Since the collision rates are expected to be very small, large detectors with low backgrounds and excellent detection capability for rare collisions are needed to detect
dark matter.


There are many direct detection experiments deployed in underground laboratories around the world.
An incomplete recent review describing these experiments can be found in Ref.~\cite{Akimov:2011za}.
When WIMPs scatter with atoms in a detection medium, they will recoil and generate kinetic motion
of atoms (heat), ionization (free electrons) and scintillation (de-excitation of excited electrons).
Direct detection experiments measure one or two or even possibly three of these signatures.
Clearly, the convenience of measuring any of these signals depends on the choice
of material. In pure semiconductors, such as those used by the CDMS collaboration~\cite{CDMS} and others, it is easy
to measure the electric current generated by electrons as well as hole carriers. In the case of noble liquid detectors
(XENON100~\cite{XENON100}, LUX~\cite{LUX}, and others \cite{others}), a light signal is usually measured by photo multiplier tubes (PMTs); ionization electrons drifting in an external electric field are either detected through their charge or through electroluminescence. In crystals, the light intensity is
usually the only signal measured. In fact, the well-known DAMA/LIBRA experiment measured  scintillation
light only~\cite{DAMA}. For heat measurements, the detector has to be kept at very low temperature, typically at tens of milli Kelvin, which is a cryogenic challenge, particularly for large masses.

Among all the direct detection experiments, the xenon dual-phase technology appears to be particularly promising. In fact, for the last 3--4 years, the XENON100 and LUX experiments
have produced the best limits over a wide range of WIMP masses~\cite{XENON100,LUX}.
There are many reasons why liquid xenon (LXe) appears to be a good choice. First,
the detection of both prompt scintillation
and ionization electrons in dual-phase mode allows not only discrimination  between nuclear and electron
recoils, but also a fiducialization of events through the time projection chamber (TPC) technology. Second, xenon does not have long-lived radioactive isotopes and can be highly purified. Third, xenon has a large atomic mass, which entails a large
WIMP scattering cross section. WIMP-nucleus scattering is coherent and hence the cross section is proportional
to the square of the atomic mass number $A$. Moreover, xenon has a large $Z$ which improves self-shielding from external
gamma rays. Finally, xenon is not prohibitively expensive, allowing detector target masses to reach ton-scale
with reasonable cost.  Xenon liquefaction temperature is around $-100^\circ$\,C, thus cryogenics is
relatively easy to manage.

The Particle and Astrophysical Xenon (PandaX) collaboration was established in 2009 and was first supported by
the Ministry of Science and Technology in China through a 973-project and by the Ministry of Education
through a 985III-project at Shanghai Jiao Tong University (SJTU). The initial
collaboration consisted mainly of physicists from SJTU, from Shandong University and
from the Shanghai Institute of Applied Physics, Chinese Academy of Sciences. The collaboration was joined
later by groups from University of Maryland, Peking University, and University of Michigan in 2011, and by the Yalong River Hydropower Development Co. Ltd. in 2012. The collaboration has been further supported by the National Science Foundation of China (NSFC), and by some of the collaborating institutions.

The PandaX experiment uses a liquid-xenon detector system suitable for both
direct dark-matter detection and $^{136}$Xe double-beta decay search. Similar to the XENON and LUX experiments,
the PandaX detector operates in dual-phase mode, allowing detection of both prompt scintillation and
ionizations through proportional scintillation. The central time projection chamber will be staged, with
the first stage accommodating a target mass of about 120\,kg, similar to that of
XENON100. In stage II, the target mass will be increased to about 0.5\,ton. In the final stage, the detector can be upgraded to a multi-ton target mass. Most sub-systems and the stage-I TPC were developed in the particle physics lab at SJTU, and have been
transported to the China Jinping underground lab (CJPL) in August 2012. After  successful installation, two engineering runs
were carried out in 2013. The system entered commissioning in December 2013
and has been collecting science data since late March 2014. A small prototype for PandaX was
developed and is running in the particle physics lab at SJTU~\cite{karl}.

The underground lab, CJPL, emerged from a government-led project to construct two large hydropower plants next to and through the
Jinping Mountain, Sichuan, China~\cite{jinping}. Jinping is located about 500\,km southwest of Chengdu, the capital of Sichuan
province. It can be accessed either by car from Chengdu, or by a short flight to Xichang, followed by a 1.5\,hr car
ride. A detailed description of CJPL can be found in Ref.~\cite{CJPL}.
Some discussion about the lab will be given in Sec. \ref{sec:background}.

In this paper, we describe the goals and the technical realization of the PandaX detector system.
In Sec. II, we describe the cryogenic system to liquefy, purify and maintain a ton-scale xenon detector.
In Sec. III, we consider the design and construction of the stage-I TPC. In Sec. IV, we discuss the properties and performance of the photomultiplier tube system. In Sec. V, we discuss the sources of background for the experiment, including cosmic rays, the shield for environmental neutrons and gamma rays, and the xenon distillation system. We also briefly discuss the background simulations. In Sec. VI, we describe the data taking electronics and data acquisition system (DAQ). We conclude the report with Sec. VII.

\section{Cryogenic and Gas-handling System}
\label{sec:cryogenics}

A reliable cryogenic and gas handling system is crucial for any noble-liquid detector system.
As the size of the detector increases, cryogenic stability, safety,  and efficiency of gas handling and purification become increasingly important issues.
The PandaX goal is to build a system storing, circulating and cooling  xenon gas up to 2--3\,tons.
So far the system has been set up and tested with ~500\,kg xenon.
A detailed description of the cryogenics system has been published in Ref.~\cite{Li:2012zs}.

\begin{figure*}[!ht]
\begin{center}
\includegraphics[width=0.49\textwidth]{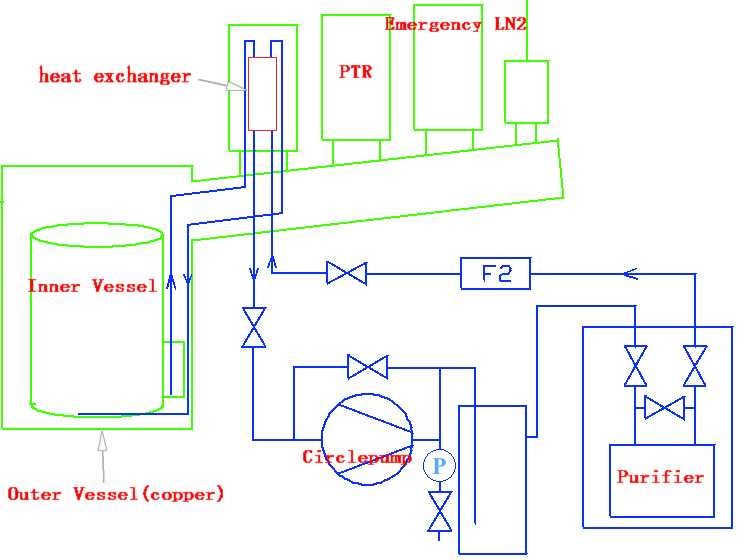}
\includegraphics[width=0.49\textwidth]{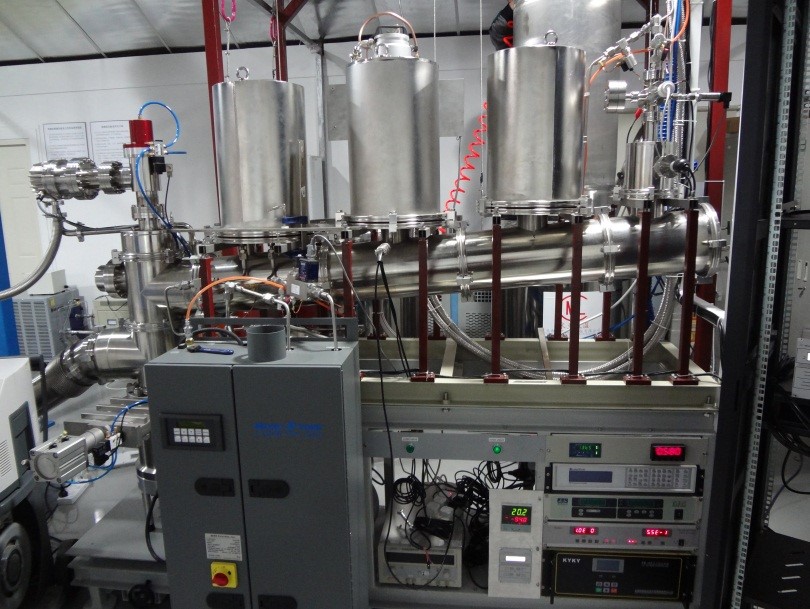}
\end{center}
\caption{Left panel: Schematic layout of the cryogenic and the gas system for PandaX. Right panel:  Photograph of the Cooling Bus for liquefaction and recirculation. In the foreground is the getter for purification.}
\label{fig:cryosys}
\end{figure*}

A schematic sketch of the cryogenic and gas handling systems is shown in Fig.~\ref{fig:cryosys} (left panel).
The detector is encapsulated in a double-walled cryostat for thermal insulation.
Additionally, seven layers of aluminized Mylar foil are used in the insulating vacuum to reduce the heat load into the cryogenic detector. The inner vessel is 0.75\,m in diameter and 1.25\,m high.
It is over-dimensioned for stage I to accommodate the future upgrade to stage II. It contains appropriate filler material to reduce the amount of xenon needed during operation. The outer vessel is made of 5\,cm high-purity oxygen-free copper.

The cryogenic system was designed as a series of independent modules, each
with a specific function. The modules are connected to the detector via common
tubing used by all modules not unlike the bus structure in a computer system
connecting all devices. This modular cryogenic system was named the Cooling
Bus (Figure~\ref{fig:cryosys},right).
Each module performs a separate function such as evacuation, heat exchange, condensation, emergency cooling and sensor mounting.
The gas handling system consists of a gas storage and recovery system, a circulation pump, a purification getter and a heat exchanger.

The detector and the Cooling Bus are connected by  concentric tubes on the evacuation module consisting of a
fore-pump and a turbo pump for the inner vessel, and a pair of pumps and a zeolite filled cryo pump for the outer vessel.
Normally the vacuum jacket is maintained by the fore-pump and the turbo pump for the outer vessel.
A cryo-sorption pump is also connected to continue
pumping the outer vacuum in case of power failures.


The system is cooled by a single Iwatani PC150 pulse tube refrigerator (PTR), driven by a 7.3\,kW air-cooled M600 compressor from Oxford Instruments.
Its cooling power was measured to be around 180\,W~\cite{Li:2012zs}.
It is mounted on a cylindrical copper block made of the oxygen-free copper.
The copper block closes off the inner chamber and  acts as a cold-finger for liquefying the xenon gas.
The PTR can thus be serviced or replaced without exposing the detector volume to air.
A copper cup that serves as a heater is installed between the PTR cold-head and the cold-finger. The temperatures of the cold-head and
cold-finger are measured by Pt100 temperature sensors.
A PID temperature controller (Lakeshore 340) regulates the heating power to keep the temperature of the cold-finger constant.

\begin{figure*}[!ht]
\begin{center}
\includegraphics[width=0.90\textwidth]{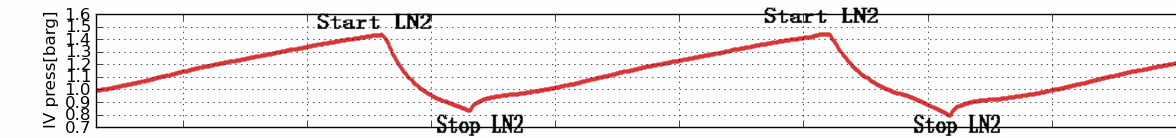}
\end{center}
\caption{Results of an emergency cooling system test: Inner pressure as a function of time when the electric power is off.}
\label{fig:ecool}
\end{figure*}

In case of failure of the PTR system, e.g. power failure, a pressure sensor will start the flow of LN2 through a cooling coil above a pressure set point about 0.5 bar above the normal operating pressure. This cooling will continue until the xenon pressure dropped below a second set point, about 0.5 bar below the normal operating pressure. The pressure sensor and the LN2 control valve are powered by a dedicated uninterruptible power supply (UPS). The warming up and LN2 cooling are shown by their effect on the xenon pressure in Fig.~\ref{fig:ecool}. Note that the emergency cooling system is always operational, but normally does not contribute to the detector cooling since the set points are never reached. This cooling can continue indefinitely if sufficient LN2 is available. As ultimate safety device a burst disc limits the maximum pressure in the detector. The burst pressure with 3.5 barA is sufficiently above the emergency set point not to trigger when LN2 cooling is available. The pressure, however, is low compared with the maximum allowable pressure of the PMT. Since the burst pressure is rather critical for safeguarding the xenon and the PMT, a
burst disc with a precisely controlled pressure rating of $\pm 5$ \% was chosen.

A heat-exchange module is used as part of the purification system which is designed to purify the liquid xenon continuously.
Constantly recirculating the xenon through a high temperature getter (SAES, PS4-M750-R-2, Max: 100\,SLPM)
gradually improves  purity
and removes electro-negative molecules originating from out-gassing of the surfaces of the detector materials.
The impurity level in LXe determines the attenuation length for scintillation light and the life time
for drifting charges. The most common electronegative impurities are  H$_2$O, O$_2$, CO$_2$, and CO).
The gas recirculation system is driven by a double diaphragm pump (KNF, PM26937-1400.12) and/or
a custom-made Q-drive~\cite{wang}. The two pumps are installed in parallel, and can backup each in cases of falling.
The flow rate in the current operation is about 30\,SLPM. The setup is sufficiently powerful to purify xenon gas in a ton-scale detector.

For ton-scale operation, hundreds of thousand of liters of xenon gas has to be recovered and stored at room temperature.
At Jinping lab, custom-made 220\,L steel high-pressure bottles and LN$_2$ dewars are used for this purpose.
Their working pressure is 8\,MPa, with each dewar storing about 250\,kg xenon gas.
These bottles can be cooled down by filling LN$_2$ into the dewars to recover xenon gas from the detector.
Tests showed that it takes 2--3 days to recover about 500\,kg xenon in the detector for the \mbox{stage-I} experiment.

To control the LXe level in the inner vessel, a volume of about 10L, called the overflow chamber, is used. The LXe can flow through a pipe from
the TPC to the overflow chamber. A Bowden Cable is attached to the end of the pipe, and then we can tune the height of the pipe's outlet from outside. With this method, the liquid level in TPC could be controlled with precision of ~0.1mm level. The liquid flowing to the overflow chamber will be recirculated through the getter and liquefied back into the inner vessel.

\begin{figure*}[!ht]
\begin{center}
\includegraphics[width=0.95\textwidth]{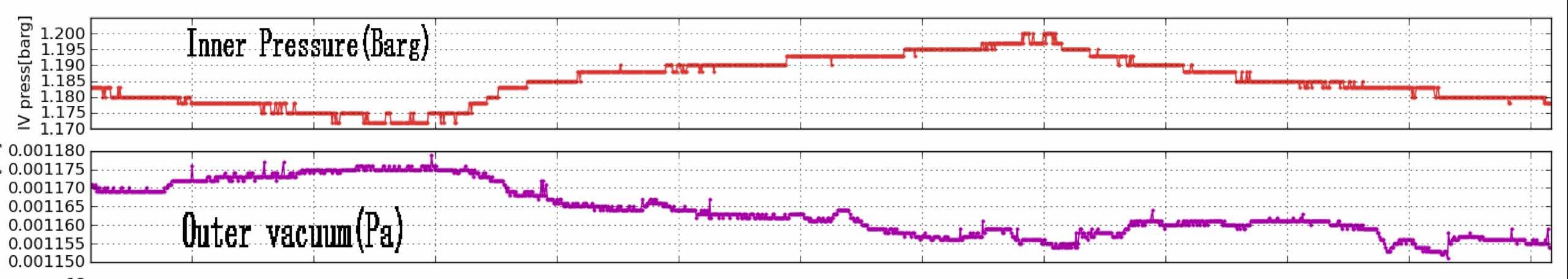}
\includegraphics[width=0.95\textwidth]{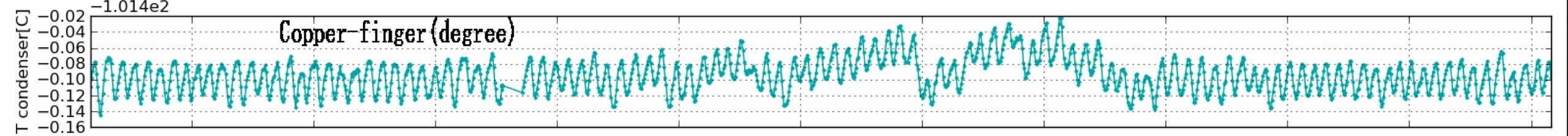}
\end{center}
\caption{Inner pressure, outer vacuum level and temperature of the cold-finger vs time, with 450\,kg liquid xenon in the inner vessel.}
\label{fig:running}
\end{figure*}

The cryogenic system for the PandaX LXe detector was tested extensively in the SJTU particle physics lab in 2012. Since moving it to CJPL in August 2012, two engineering runs were performed. The system was filled with
450\,kg of xenon, which was later recovered. It takes approximately 3\,days to liquefy the xenon gas with LN$_2$ assisted cooling.
Each time the system performed as expected.
Fig.~\ref{fig:running} shows  typical values  for inner pressure, outer vacuum level, and cold finger temperature as a function of time
as the system is running. These values can be accessed through a slow-control system via the internet. They indicate stable running conditions.

\section{PandaX Stage-I TPC}
\label{sec:tpc}

The time projection chamber for the PandaX stage-I experiment is optimized for light collection efficiency to achieve a low energy threshold
for a high sensitivity to light dark matter at around 10\,GeV/c$^2$. The diameter of the field cage is 60\,cm and the drift length is 15\,cm, accommodating a liquid xenon target mass of 120~kg.
For stage II, the design is similar except for the drift length which will be increased to 60\,cm.

\begin{figure*}[!ht]
\begin{center}
\includegraphics[width=0.40\textwidth]{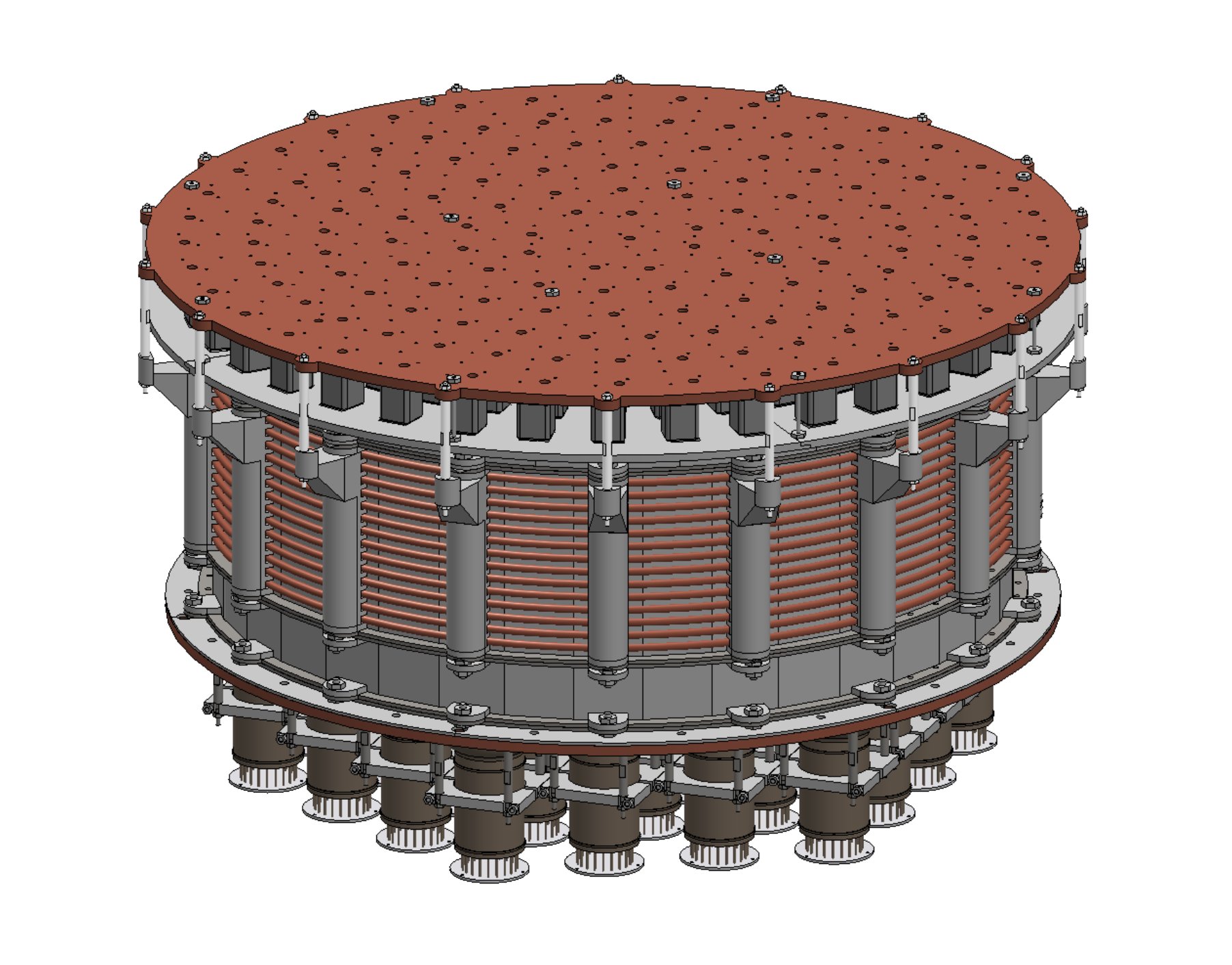}
\includegraphics[width=0.59\textwidth]{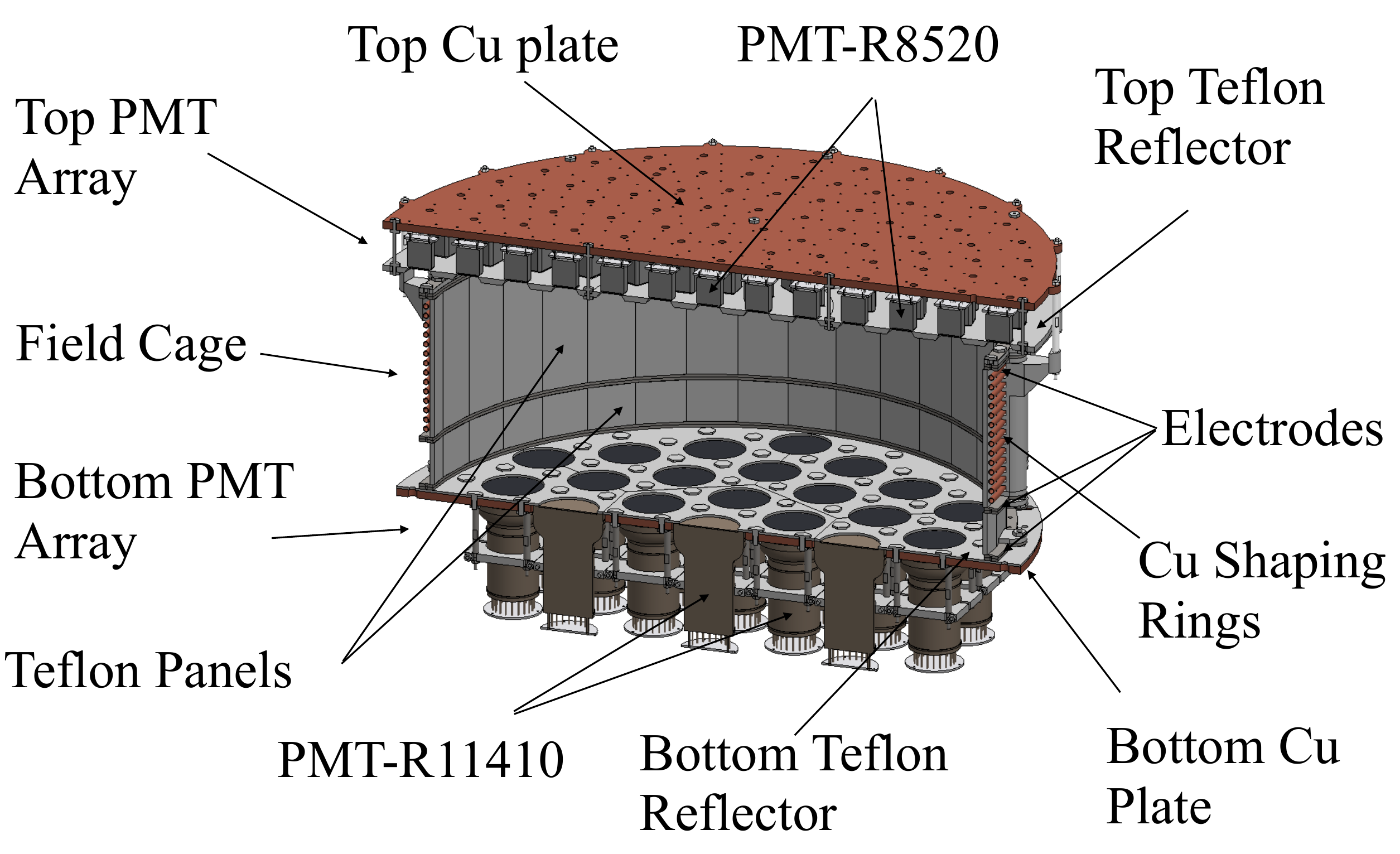}
\end{center}
\caption{Mechanical design of stage-I TPC showing the top PMT array, the field cage and the bottom PMT array. Left panel: Full view showing the completely integrated TPC. Right panel: Cross-sectional view showing the detailed components.}
\label{fig:design-stage1a}
\end{figure*}

The mechanical design for the TPC is shown in Fig.~\ref{fig:design-stage1a}. There are three main parts to the TPC: the top PMT array, the field cage and the bottom PMT array. In the following subsections, we will discuss
in detail its construction.

\subsection{Field cage}

\begin{figure*}[!ht]
\begin{center}
\includegraphics[height=1.45in]{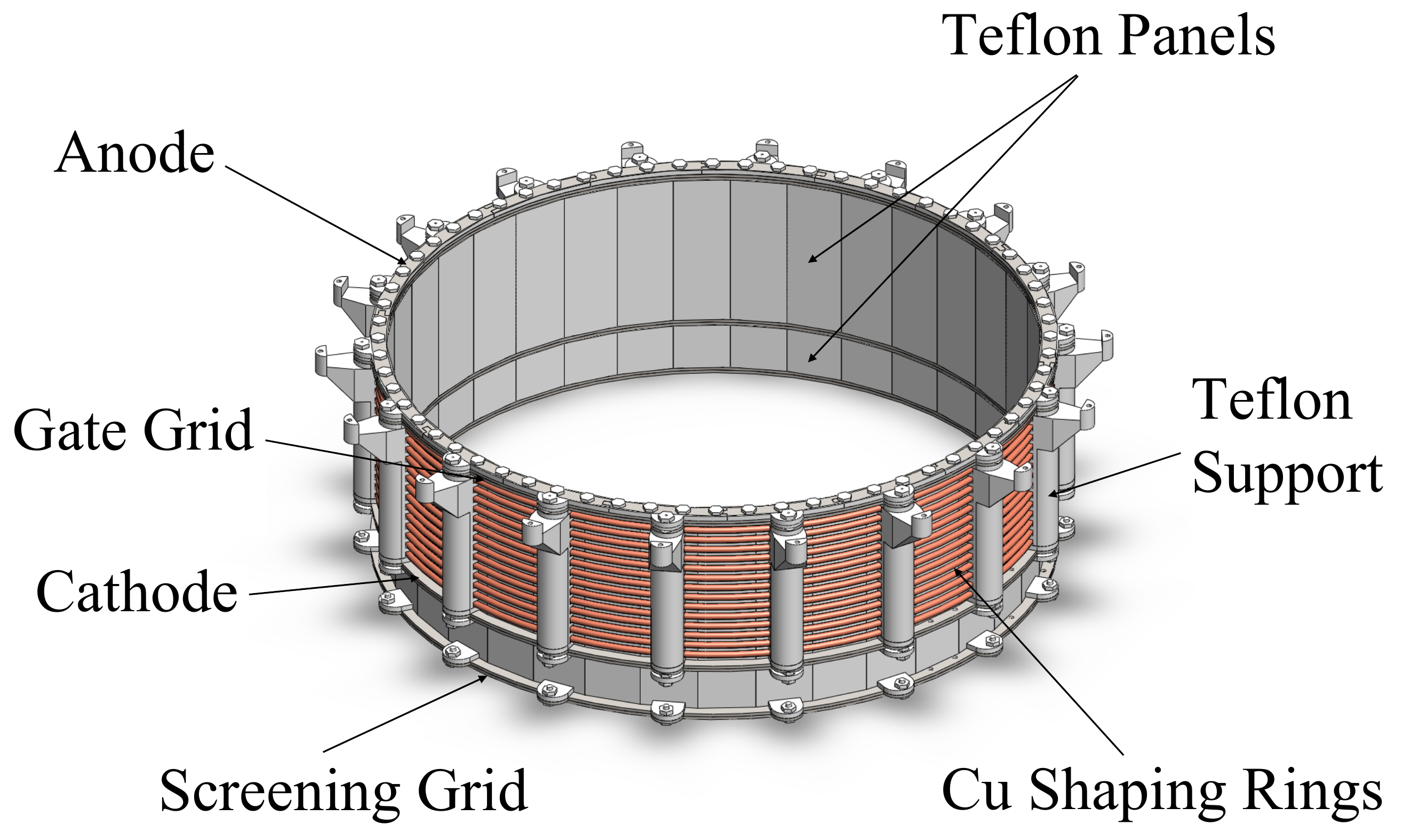}\hspace*{2mm}
\includegraphics[height=1.45in]{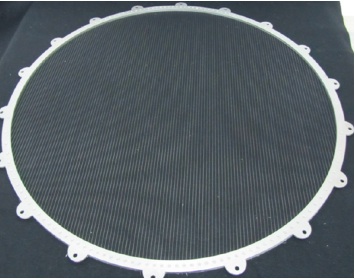}\hspace*{2mm}
\includegraphics[height=1.45in]{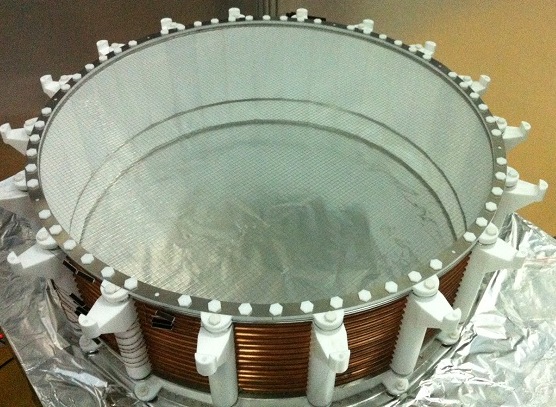}
\end{center}
\caption{Left panel: Mechanical design of the field cage showing the teflon panels and supports, three electrodes and the field shaping rings. Center panel:  Photograph of the wire electrode for the gate grid, the cathode and the screening electrodes. Right panel:  Photograph showing the anode mesh is fixed above the grid ring with a teflon ring in between.}
\label{fig:electrodes}
\end{figure*}

The field cage, shown in Fig.~\ref{fig:electrodes}, which is the main part of the TPC, contains a teflon cylinder of 15\,cm height and 60\,cm diameter.  The teflon cylinder has 36 pieces of 5-mm thick teflon panels interlocking with adjacent panels. Among them, 18 pieces of teflon panels are designed with 5-mm thick tablets on their top and bottom ends, by which each of them are fixed on a teflon support by teflon bolts. Those 18 pieces of teflon supports serve as the bearing carrier of the field cage and hold all other components besides the teflon panels. There are altogether four electrodes named anode, gate grid, cathode and screening electrodes from top to bottom in the TPC. Each electrode has an inner diameter of 60\,cm, which is the same as that of the field cage.


To fabricate the gate grid, the cathode and the screening electrodes, stainless steel wires (304 or 316L) with 200\,$\mu$m diameter are pressed between two stainless steel rings by screws (see Fig.~\ref{fig:electrodes} center). The rings are made of  316L stainless steel with 3\,mm thickness and 15\,mm width. The spacing between two wires is 5\,mm, which results in 96\% optical transparency for the electrode.  Each wire has a tension of 2.8\,N (43\% of yield strength), provided by 288\,g weight rods hanging on both side of each wire during production. The anode electrode is made of photo-etched mesh with crossing bars of 200\,$\mu$m width and 5\,mm spacing, providing 92\% optical transparency. The anode mesh is fixed above the grid with 5-mm thick teflon rings (see Fig.~\ref{fig:electrodes} right). Thus the distance between the anode mesh and the gate grid wires is 8\,mm.

\begin{figure*}[!ht]
\begin{center}
\includegraphics[height=1.5in]{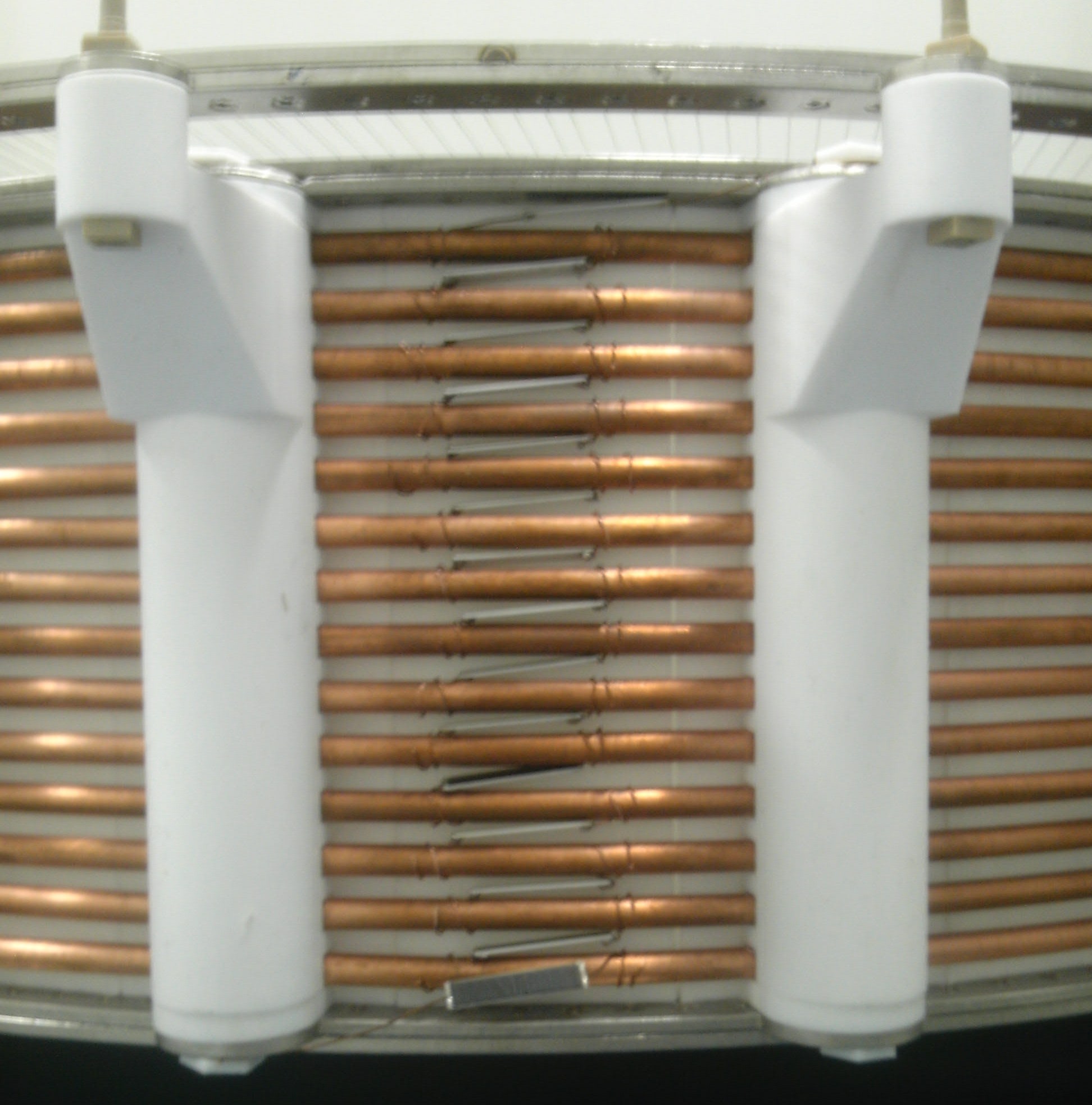}\hspace*{3mm}
\includegraphics[height=1.5in]{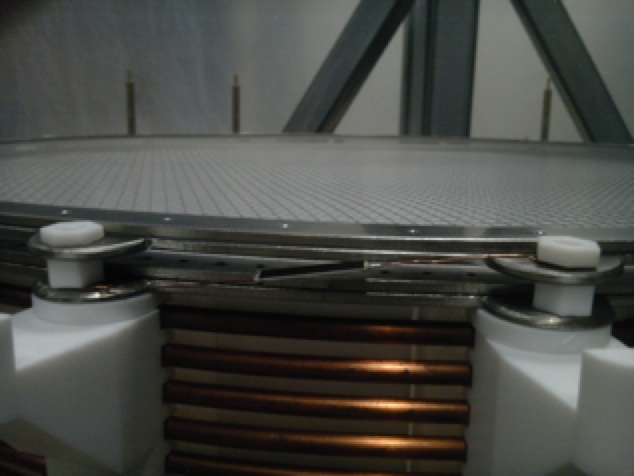}\hspace*{3mm}
\includegraphics[width=0.26\textwidth]{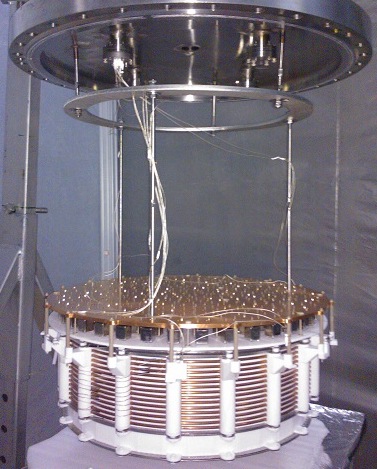}
\end{center}
\caption{Photographs showing the resistors between the shaping rings, the gate grid and the cathode (left), a resistor mounted between anode and the gate grid (center), and the field cage, together with the top PMT array, hanging from the top flange of the inner vessel (right).}
\label{fig:resistor}
\end{figure*}

The anode, the gate grid and the cathode electrodes are fixed on the teflon supports as shown in Fig.~\ref{fig:electrodes}. The teflon supports are mounted to the top copper plate by PEEK bolts so the field cage is integrated with the top PMT array. To make the drift field uniform, 14 pieces of OFHC copper shaping rings were arranged outside the teflon panels, between the gate grid and the cathode in equal spacing. They are made of OFHC copper tubes with 6\,mm outer diameter and 5\,mm inner diameter, and clamped by teflon supports and teflon panels (see Fig.~\ref{fig:resistor}). 500\,M$\Omega$\, surface-mount resistors (SM20D from Japan FineChem Company, Inc.), rated for maximum voltage of 5 kV, are tied by bare copper wires between each two adjacent shaping rings, between shaping rings and cathode or gate grid, and between gate grid and anode (see Fig.~\ref{fig:resistor}). Two resistor chains are mounted on the electrodes to prevent interruption of the experiment if a single resistor breaks during operation. A photograph showing the field cage, integrated with the top PMT array, installed in the detector is shown in Fig.~\ref{fig:resistor}.

\subsection{PMT arrays}


The bearing carrier of the top PMT array is an 8-mm thick oxygen-free high-conductive (OFHC) copper plate which holds 143 pieces of Hamamatsu R8520-406 PMTs.
All top PMTs are fixed on bases by pin-socket connections directly while all bases are mounted on the OFHC copper plate by spring-loaded screws. From the edge to the center, the 143 PMTs are arranged uniformly in 6 concentric rings around a center PMT. The spacing of adjacent rings is 52.5\,mm and the number of PMTs in each ring is 36, 36, 28, 20, 14 and 8 from the edge to the center. To achieve a better position reconstruction for events at large radius, the diameter of the outermost PMT ring is 630\,mm, which is larger than that of the field cage. The teflon reflector covering the space between PMTs is made of five pieces of 6-mm thick teflon plates with openings according to the arrangement of PMTs. Those teflon plates are ``interlocking'' with each other to avoid gaps from appearing due to the shrinkage of teflon at LXe temperature, and are mounted on the OFHC copper plate by PEEK bolts.

\begin{figure*}[!ht]
\begin{center}
\includegraphics[height=1.55in]{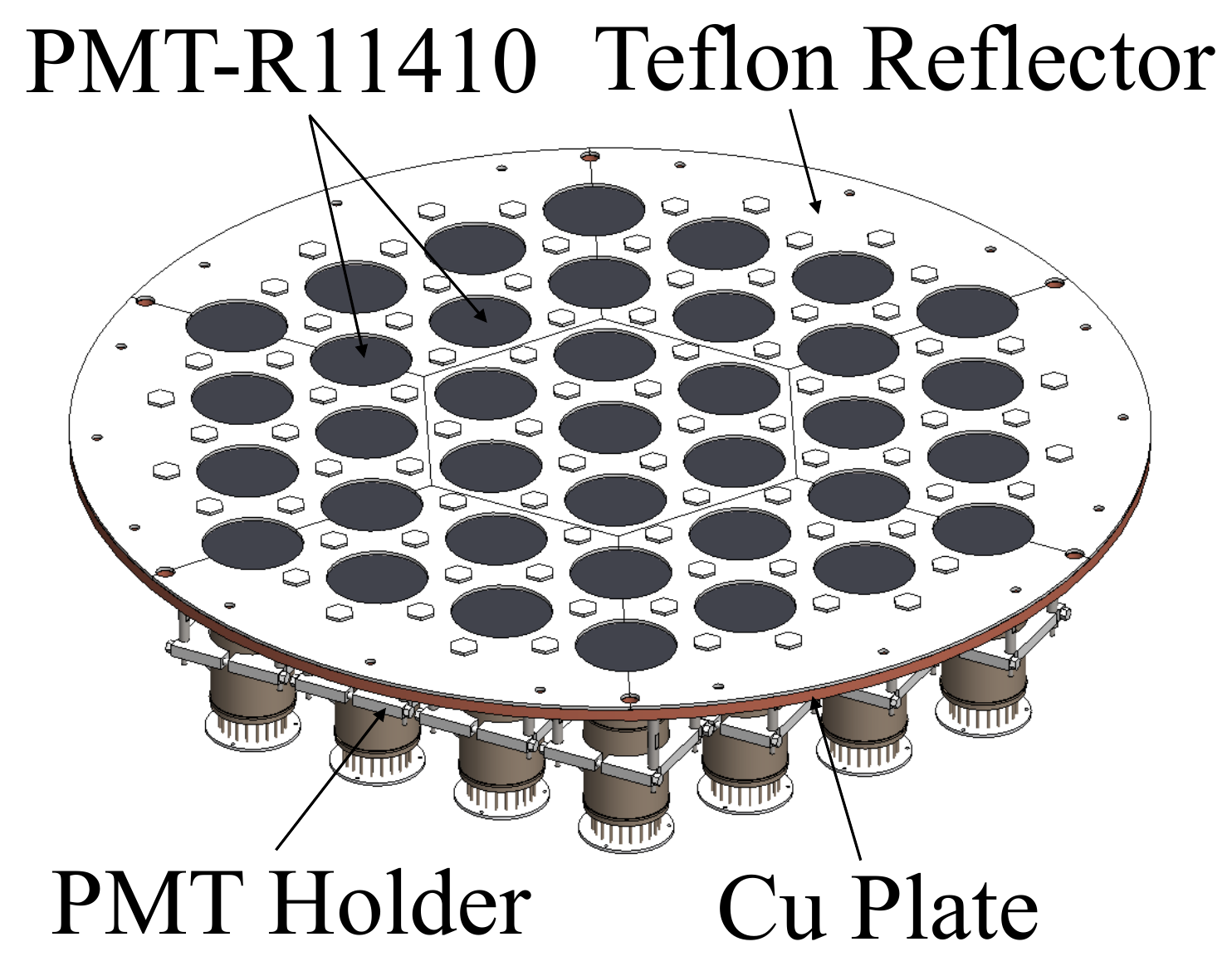}\hspace*{3mm}
\includegraphics[height=1.55in]{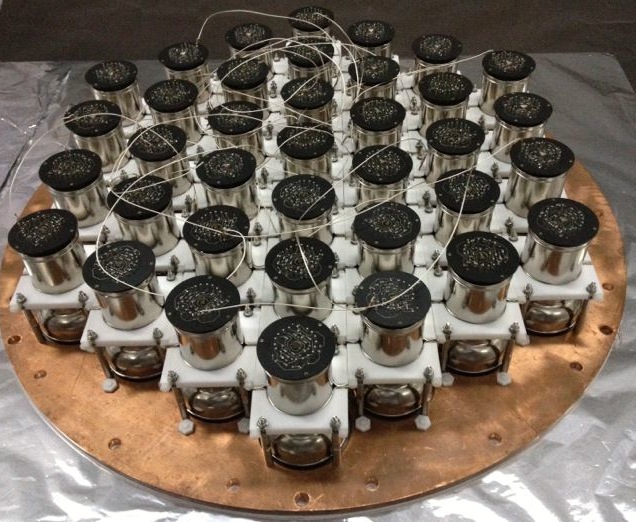}\hspace*{3mm}
\includegraphics[height=1.55in]{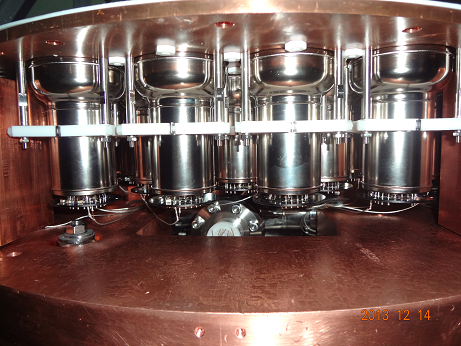}
\end{center}
\caption{Left panel: Mechanical design of the bottom PMT array and the teflon reflectors. Center panel: Photograph showing all 37 R11410 PMTs installed on the copper plate of the bottom PMT array. Right panel: Photograph showing the bottom PMTs with the copper filler and how they are connected to the bottom feedthroughs.}
\label{fig:botpmtpic}
\end{figure*}

The bottom PMT array is supported by a 8\,mm thick OFHC copper plate which holds the 37 pieces of Hamamatsu R11410-MOD PMTs. Unlike the top PMTs, each bottom PMT is held by a pair of stainless steel clamps, fixed in between two 8\,mm thick teflon clamps which are fixed on the OFHC copper plate by stainless steel bolts, and all bases are fixed on each PMT by pin-socket connection directly, as shown in Fig.~\ref{fig:botpmtpic}. Teflon reflectors are mounted  between the PMT windows to increase light collection. The teflon reflector is made up of 7 pieces of 2.4-mm thick teflon sheets, which are fixed on the OFHC copper plate by teflon bolts. To reduce the electric field strength near the bottom PMT photocathode to acceptable levels, the grounded screening electrode wires are at 5\,mm above bottom PMT windows. Similar to in the field cage, 36 pieces of shorter teflon panels are designed to improve light collection efficiency, and fixed on the OFHC copper plate by sharing the same teflon bolts with the screening electrode. The entire array with all PMTs assembled on the copper plate is shown in Fig.~\ref{fig:botpmtpic}. The PMT HV and signal cables are connected on the bottom flange of the inner vessel. To minimize the empty space below the PMTs and to shield gamma rays from the bottom flange, a copper filler is installed below the PMT array (see Fig.~\ref{fig:botpmtpic}).


\subsection{Initial operation of the TPC}

For a dual phase operation of the PandaX TPC, the anode is connected to the ground, while the gate and cathode grids are connected with negative high voltages (HV). The cathode is connected to a custom-made HV feedthrough, while the gate grid is connected to a commercial feedthrough rated up to 10 kV. During the engineering run in liquid xenon with no PMT signal readout, the high voltage on the cathode reached 36\,kV before electrical breakdown, and the high voltage on the gate grid reached 6\,kV (the maximum value of the power supply). However, during the full operation of the detector with PMT readout, micro-discharge signals were observed by the PMTs when the cathode voltage reaches 20~kV. Similar discharge signals were observed when the gate grid is above 5~kV. These discharges produce many small signals at the single or a few photoelectrons level, preventing the useful data taking. Thus it sets the limitation of the drift field at 1~kV/cm across the 15~cm drift gap.

To extract electrons from the liquid to the gas phase, we set the liquid level in between the anode and the gate grids. The liquid level can be adjusted with an overflow point controlled by an external motion-feedthrough. The level should be set at least 3~mm above the gate wires so that it covers the 3~mm-thick stainless steel ring to avoid discharges from any sharp point on the ring if it's exposed in the gas. We set the level at 4~mm above the gate grid for the initial operation of the TPC, giving a 4~mm gas gap.  For the extraction field in the gas above the liquid xenon, a 5~kV voltage between the anode and gate grids provides an extraction field of 8.3~kV/cm, corresponding to an electron extraction yield of 90\% according to~\cite{Aprile:JPG14}.

Following the adjustment of high voltages on the cathode and gate grids, both S1 and S2 signals are observed by the PMTs after the liquid xenon reached a good purity.
A typical S1-S2 waveform summed from all PMTs and their S2 signal distribution among the PMTs for a single-site event is shown in Fig.~\cite{fig:waveform}.

\begin{figure*}[!ht]
\begin{center}
\includegraphics[height=2.8in]{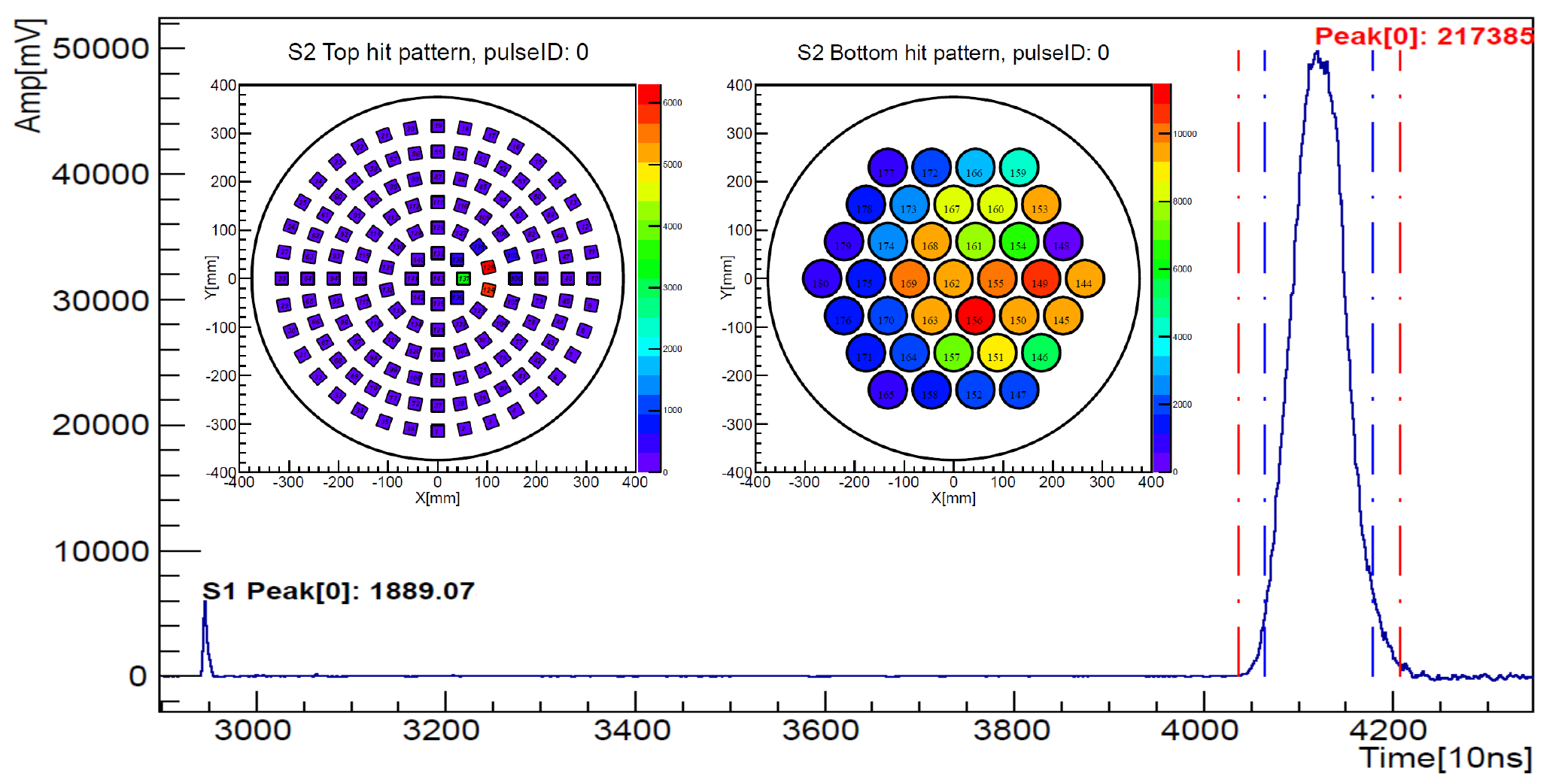}\hspace*{0mm}
\end{center}
\caption{The summed waveform and S2 signal distribution on all PMTs of a typical single-site event during the calibration run.}
\label{fig:waveform}
\end{figure*}

\section{Photomultiplier system}
\label{sec:pmt}
PandaX uses specially-developed PMTs to detect prompt and proportional scintillation light. The photomultiplier system has to satisfy many requirements, such as good quantum efficiency to VUV light (178nm) from xenon, low radioactivity, single photoelectron (SPE) resolution, good timing resolution, cryogenic operation ($-100^\circ$\,C) suitability, high-pressure operation ($>$3\,atm) suitability, and minimal outgassing from the bases and cables. These requirements are met by the one-inch Hamamatsu R8520-406 and by the three-inch Hamamatsu R11410-MOD  photomultiplier tubes, which instrument the top and
bottom photomultiplier arrays, respectively. In this section, we describe various aspects
of the PMT system, including the basic properties, bases, high voltage and decoupler, feedthrough and cabling, calibration and test results.

\subsection{Photomultiplier tubes}

The Hamamatsu model R8520-406 is a 10-stage, one-inch square photomultiplier tube,
rated for a temperature range of $-110^\circ$\,C to $+50^\circ$\,C, and 5-atm pressure resistance. The
typical gain is $10^6$ at a HV setting of 800\,V. The
cathode window is made of Synthetic Silica, and the cathode material
is Bialkali, yielding an excellent quantum efficiency of about 30\% at 175\,nm. The
radioactivity of this tube has been measured by XENON100 collaboration in Ref.~\cite{ref:Xenon_rad}.


The Hamamatsu model R11410-MOD with ceramic stem is a 12-stage, three-inch circular photomultiplier
tube, also rated
for LXe temperature  with a typical gain of $5\times10^6$ at 1500\,V.
The maximum pressure rating, updated in 2011, is
0.4\,MPa (absolute)~\cite{ref:R11410-MOD}. The quantum efficiency is $>$30\% at
175\,nm. Radiopurity measurements were performed
elsewhere~\cite{ref:Xenon_rad}, as well as in our own counting station at CJPL (see Sec.~\ref{Sec:counting-station}).
Note that for low temperature operation, flying leads were cut off to a length of
10\,mm from the ceramic stem for attachment to the voltage divider.


\subsection{Voltage dividers and bases}

Positive HV voltage dividers (bases)
are chosen to put the photocathode at ground a) to reduce
the noise level and improve the single photoelectron (SPE)
resolution, b) to eliminate potential
interference with the electric field profile of the TPC, and c) to allow
bringing signal and HV in/out the vessel through the same coaxial cable
(decouplers will be placed outside the vessel) to minimize the
number of cable feedthroughs.

\begin{figure}[!htb]
\centering
\subfigure[~R8520-406] 
{
  \label{fig:R8520_sch}
  \includegraphics[width=0.9\textwidth, angle=0]{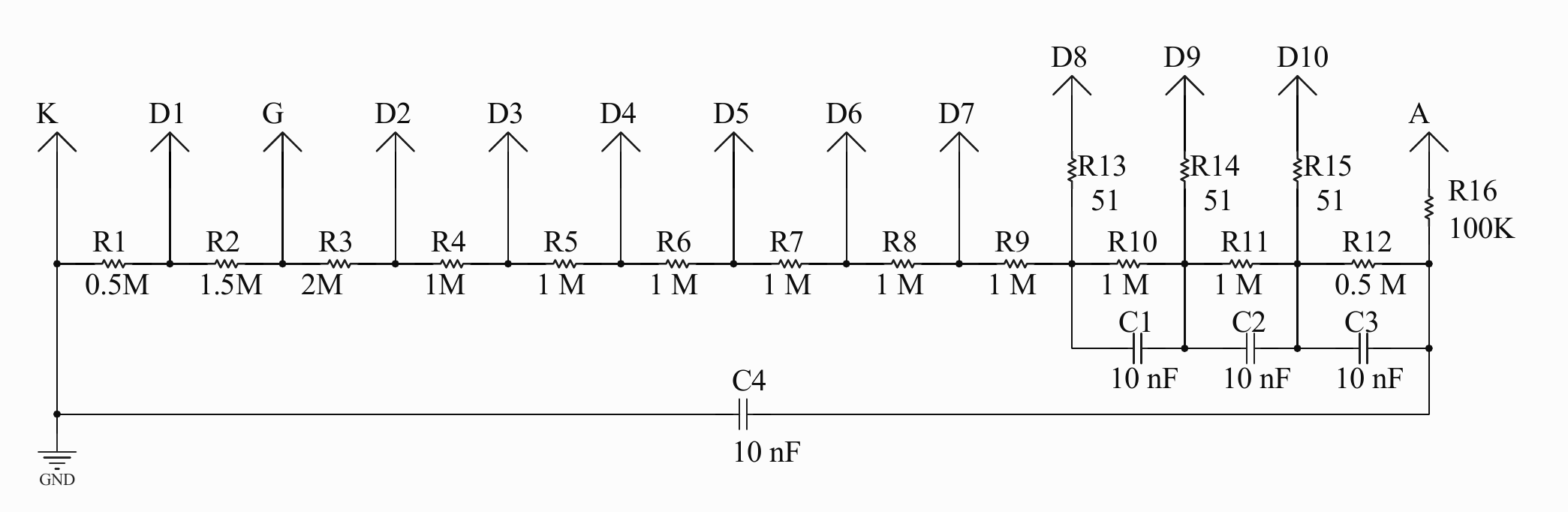}
}
\subfigure[~R11410-MOD] 
{
  \label{fig:R11410_sch}
  \includegraphics[width=0.9\textwidth, angle=0]{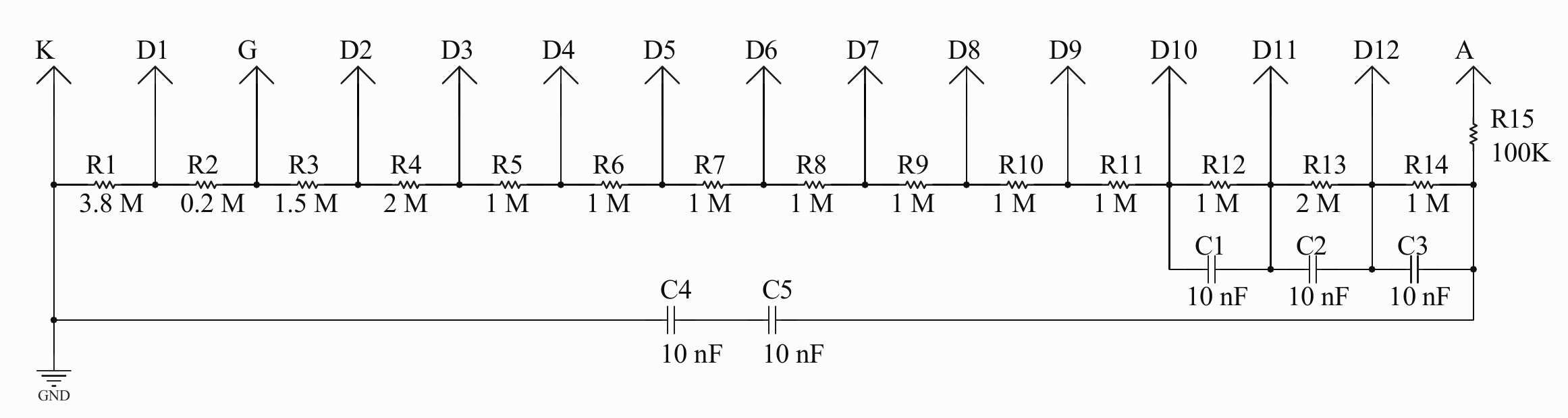}
}
\caption{PMT base schematics for (a) the R8520-406 and (b) the R11410-MOD bases.
}
\label{fig:base_sch} 
\end{figure}

The design of the voltage divider for the R8520-406 PMT follows
recommendations from Hamamatsu, with schematics
shown in Fig.~\ref{fig:R8520_sch}.
To limit the heat load for cryogenic operation, the total resistance of
the divider chain is 13\,MOhm, so that heat power from each base is only 0.05\,W
under normal voltage setting (800\,V). The base is back-terminated
with 100\,kOhm resistor R$_{16}$, which increases not only the charge
collection at the output end but also the low frequency band
width, which is critical to minimize signal distortion for S2-type  signals.
With the rest of the frontend electronics terminated at 50\,Ohm, this results in some signal reflection
(see Sec.~\ref{sec:bench}). A 10\,nF capacitor C$_{4}$
between the cathode and anode is necessary to remove signal oscillation.

The design of the R11410-MOD base is very similar to that of the R8520-406 base, with its schematic
shown in Fig.~\ref{fig:R11410_sch}. For all base capacitors,
low background ceramic X7R capacitors from Kyocera Inc., rated at 1\,kV, were selected.
Two capacitors were put in series between the cathode and the anode for the R11410-MOD base.
To reduce radioactivity, C$_{2}$ and
C$_{3}$ were removed from both bases. Signal distortion of typical S2 pulses were
measured to be negligible (see Sec.~\ref{sec:bench}).

\begin{figure}[!htbp]
\centering
\subfigure[~R8520-406] 
{
  \includegraphics[height=2.2in]{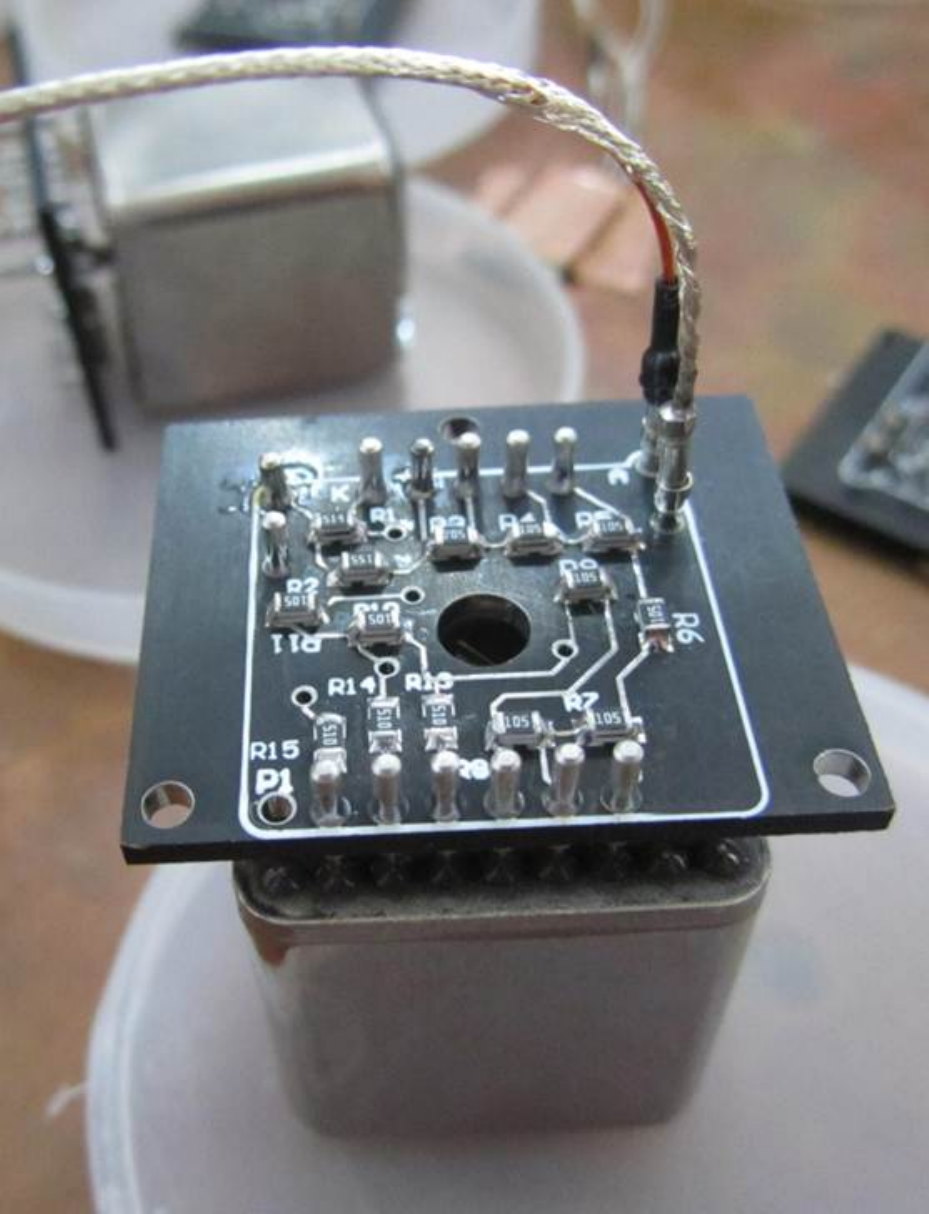}
}
\subfigure[~R11410-MOD] 
{
  \includegraphics[height=2.2in]{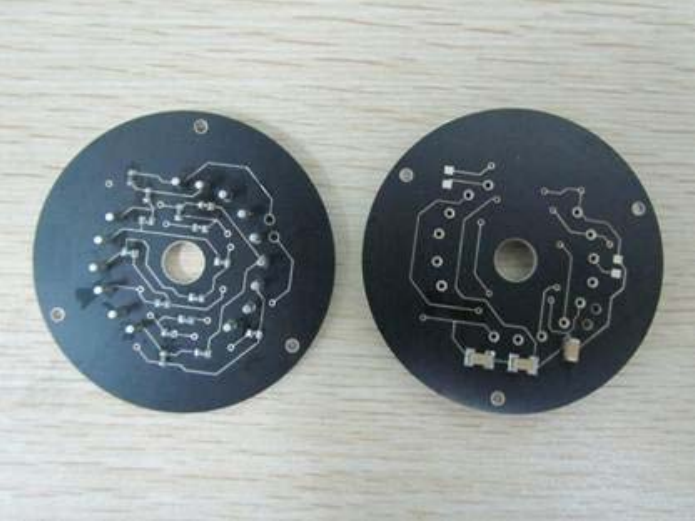}
}
\caption{Photographs of the PMT bases for  (a) the R8520-406 and (b) the R11410-MOD models.
}
\label{fig:bases}
\end{figure}

A photograph of an R8520-406 base is shown in Fig.~\ref{fig:bases}.
Several technical issues were addressed in its construction.
Cirlex, a kapton-based material, was chosen as the base material due to its good radiopurity and low outgassing characteristics. Pure silver tracks were deposited onto the
PCB without other add-ons. Ceramic capacitors from Kyocera,
and lead-free soldering tin (Sil-Fos) from Lucas-Milhaupt/Handy \& Harman
were selected and used on the base to minimize radioactivity.
KAP3, a UHV coaxial cable
from MDC Inc was selected as the signal/HV cable. The receptacles for the
PMT pins were chosen from Mil-Max Inc. The base of the R11410-MOD PMT was
constructed in a very similar way, as shown in the right panel of Fig.~\ref{fig:bases}.

The PMTs and bases were radioassayed in the HPGe counting station
in Jinping,
with results summarized in Sec.~\ref{Sec:counting-station}. The radioactivity levels
from the PMTs are in agreement with those reported in Ref.~\cite{ref:Xenon_rad}.
The high
$^{238}$U/$^{232}$Th/$^{40}$K content in the one-inch base is likely due to the
particular type of pin receptacles used, since other material components
are identical to those on the three-inch bases.

\subsection{Signal-HV decoupler and high voltage system}
\begin{figure}[!htbp]
\centering
\includegraphics[width=0.45\textwidth]{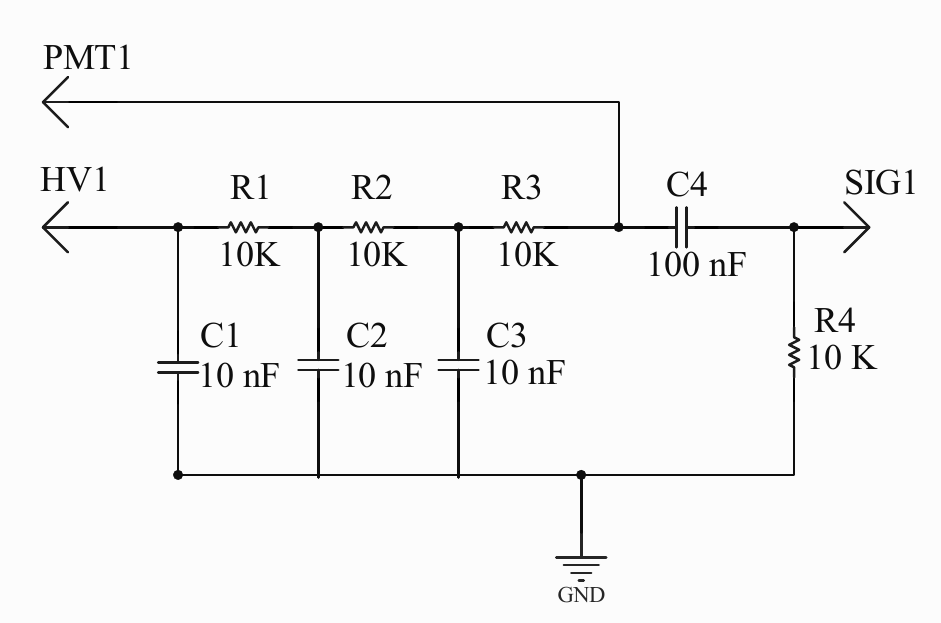}
\includegraphics[width=0.45\textwidth]{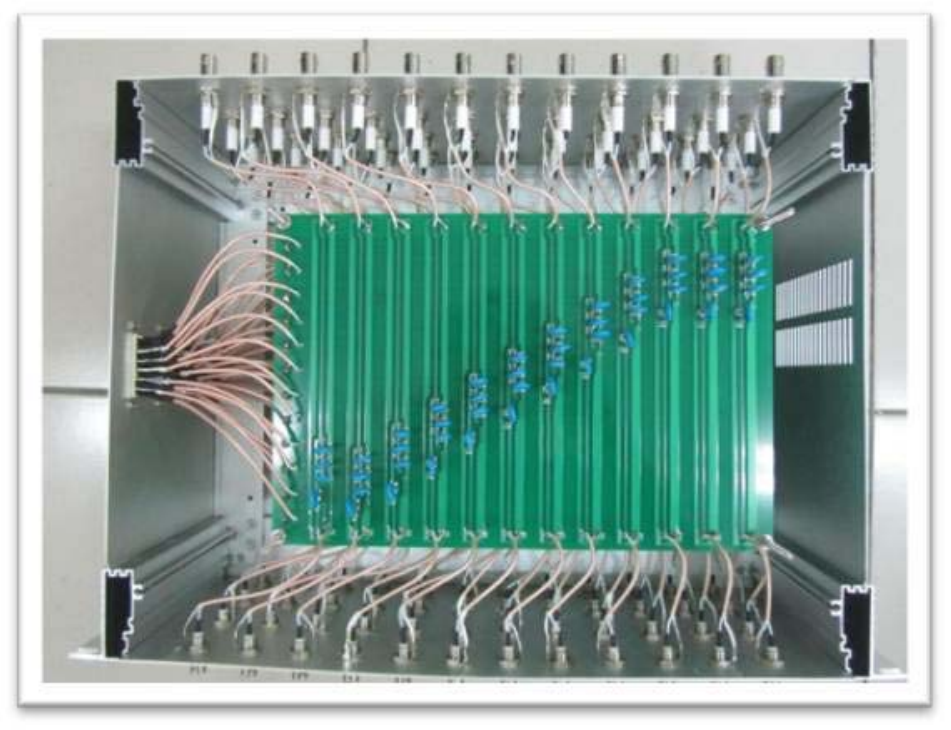}
\caption{Left panel: Schematic diagram of a single channel signal-HV decoupler circuit. Right panel: Photograph of a 48-channel decoupler module.}
\label{fig:decoupler}
\end{figure}

The signal (fast pulses) and the DC high voltage are
decoupled outside the detector via a decoupler module.
A similar design from the Daya Bay experiment~\cite{ref:decoupler} was followed.
A schematic sketch of the decoupler is shown in Fig.~\ref{fig:decoupler} (left).
The decoupling capacitor, rated for 2\,kV, is chosen to be 100\,nF based on
a SPICE simulation and bench tests to minimize the signal
distortion for S2-like signals. At the input
of the DC high voltage, a 3-stage high pass filter is implemented to remove
the ripples from the high voltage supply.

A photograph of a 48-channel decoupler module (4 U) is
shown in Fig.~\ref{fig:decoupler} (right).
Each module consists of four
12-channel PCBs. The high voltage
inputs into the module are through four China-standard mil-spec DB25 connectors, tested for 2\,kV. Cables leading to PMTs are through
48 SHV connectors on the back panel, and the decoupled signals to the electronics
are through the 48 BNC connectors on the front panel.

The PMT high voltage system is from CAEN SpA, with a
SY1527LC~\cite{ref:SY1527} main frame
and four A1932AP~\cite{ref:A1932A} modules, each supplying 48 HV channels
up to 3\,kV. The ground of these HV channels is configured as float
from the ground of the main frame. The output connector on each module is a
52-pin HV connector by Radiall. A custom fan-out cable is made
to connect each Radiall connector to four DB-25 connectors, as the
side input to the decoupler box (see Fig.~\ref{fig:decoupler}).
To reduce the 200\,kHz noise from the HV supply, additional RC filters
were implemented before the HV enters into the decoupler box.

\subsection{Cabling}
\begin{figure}[!htbp]
\centering
\includegraphics[width=0.9\textwidth]{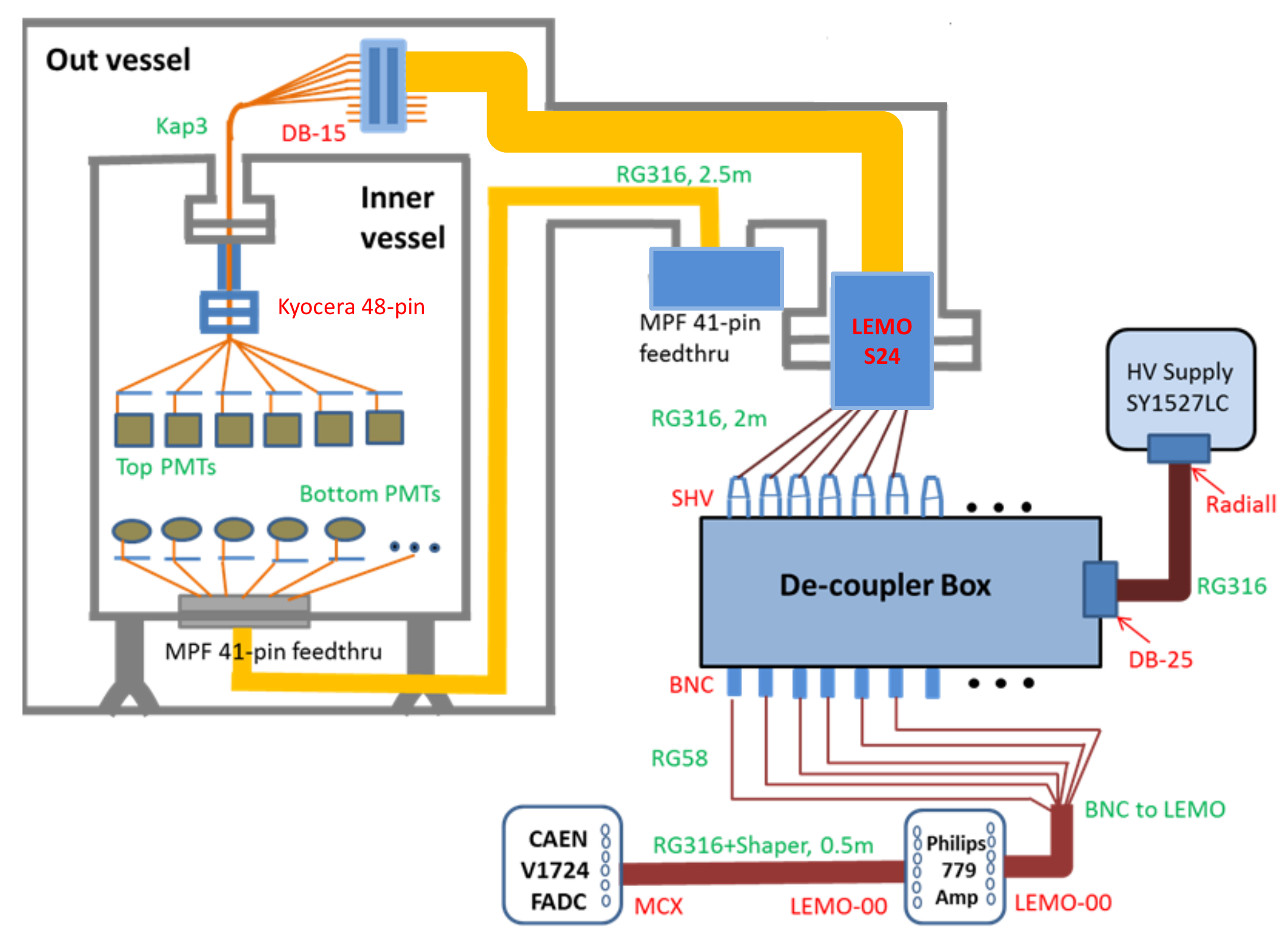}
\caption{Schematic overview of the PandaX cabling scheme.}
\label{fig:cabling}
\end{figure}

The PandaX PMT cabling scheme is shown in Fig.~\ref{fig:cabling}.
As mentioned previoulsy, the KAP3 cables connect to individual PMT bases. For the top
PMT bases, the cables exit the inner vessel through six
48-pin double-sided high voltage CF35 feedthrough flanges by Kyocera.
The lower PMT feedthroughs
are two commercial double-ended 41-pin low temperature HV feedthroughs by MPF Products Inc.
The connectors to these feedthroughs are custom-made with PEEK as insulator
and lead-free sockets by TE/AMP.

The RG316 coaxial cables carry the PMT signals/HV into the outer vessel vacuum.
From the feedthrough flange on the inner vessel, every six cables
are grouped with the other end soldered onto a male China-standard
mil-spec DB15 connector, with core and ground individually separated.
A custom cable assembly, consisting of 12 RG316 cables,
connect two of these connectors to a 24-pin double-sided vacuum feedthrough
by LEMO Inc.
Eight such LEMO feedthroughs are sealed against a ISO160 flange on the
outer vessel via o-rings. For the
bottom PMT cables, a cable assembly with RG316 cables connects
the bottom feedthrough with another MPF double-sided 41-pin feedthrough mounted
on the wall of the outer vessel.
The cables outside the outer vessels are RG316 as well, each connecting to the
decoupler modules through individual SHV connectors. The total cable length
from the decoupler to the PMT base is approximately 5\,m.

\subsection{LED calibration system}
A fiber optics system is installed in the detector to carry external
LED light pulses into the detector for single photoelectron calibration.
Optical fibers feed into the inner vessel through commercial ultra-high
vacuum fiber feedthroughs. The open ends of the fibers were inserted
into three teflon rods
mounted on the wall of the TPC.
Three external blue LEDs are driven by custom-built
pulsers~\cite{ref:LEDPulser}. Fast light pulses ($<$10\,ns) are triggered by a
TTL pulse, which also serves as the external trigger of the frontend electronics,
with its intensity controlled by a negative DC voltage.

\subsection{PMT performance during commissioning}
\label{sec:bench}
\begin{figure}[!htbp]
\centering
  \includegraphics[width=0.45\textwidth]{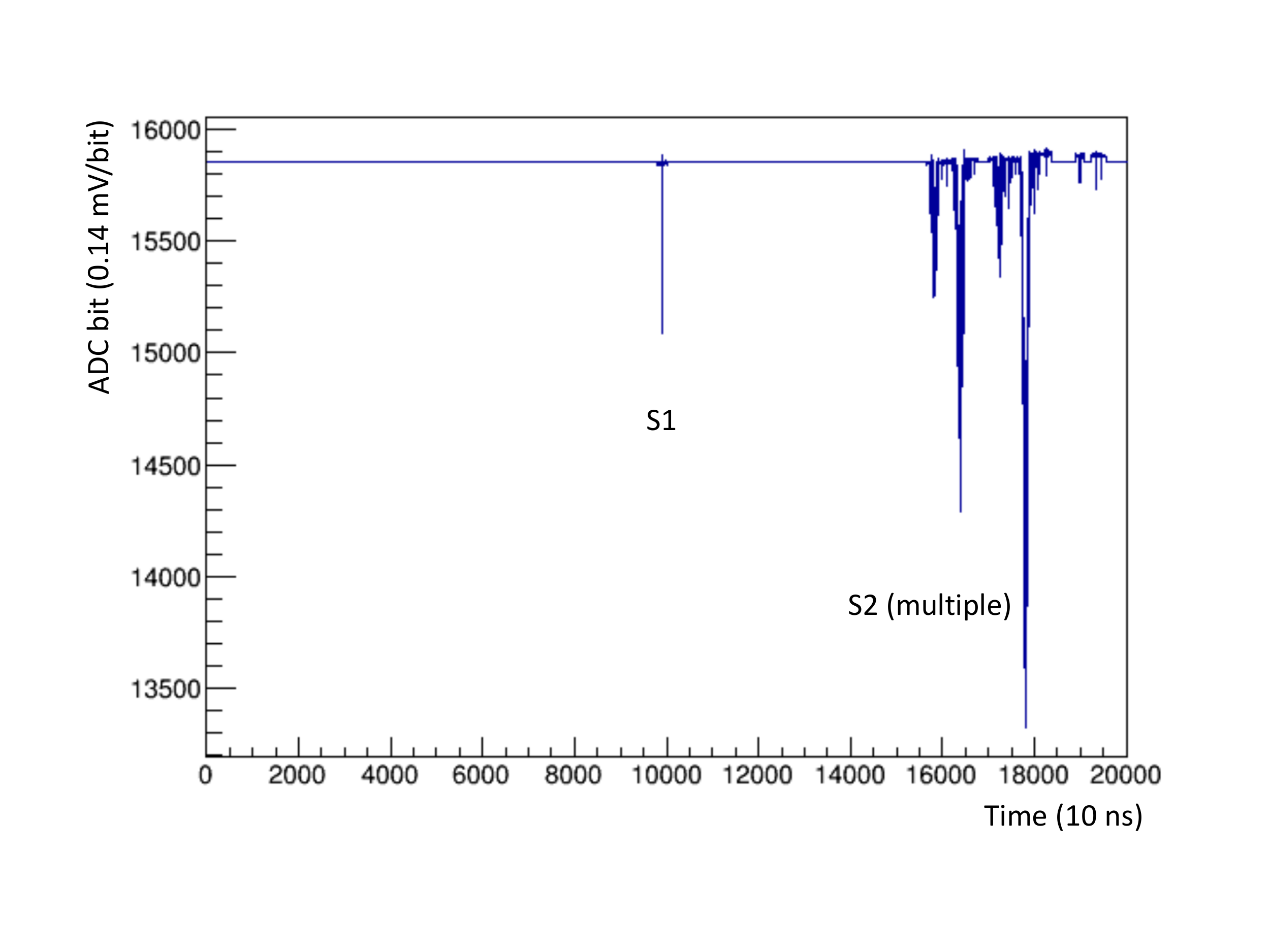}
\caption{A typical multi-site event waveform seen by a bottom PMT. The waveform is digitized by the CAEN V1724 FADC.}
\label{fig:wf} 
\end{figure}

PMT properties were measured with the full electronics chain during commissioning of the PandaX detector. Liquid xenon was filled into the
detector with the liquid-gas interface located between the grid of the extraction
field and the anode. A typical S1-S2 waveform from a bottom PMT measured by the digitizer is shown in Fig.~\ref{fig:wf}, in which a multi-step scattering is clearly identified.

\begin{figure}[!htbp]
\centering
\includegraphics[width=0.45\textwidth]{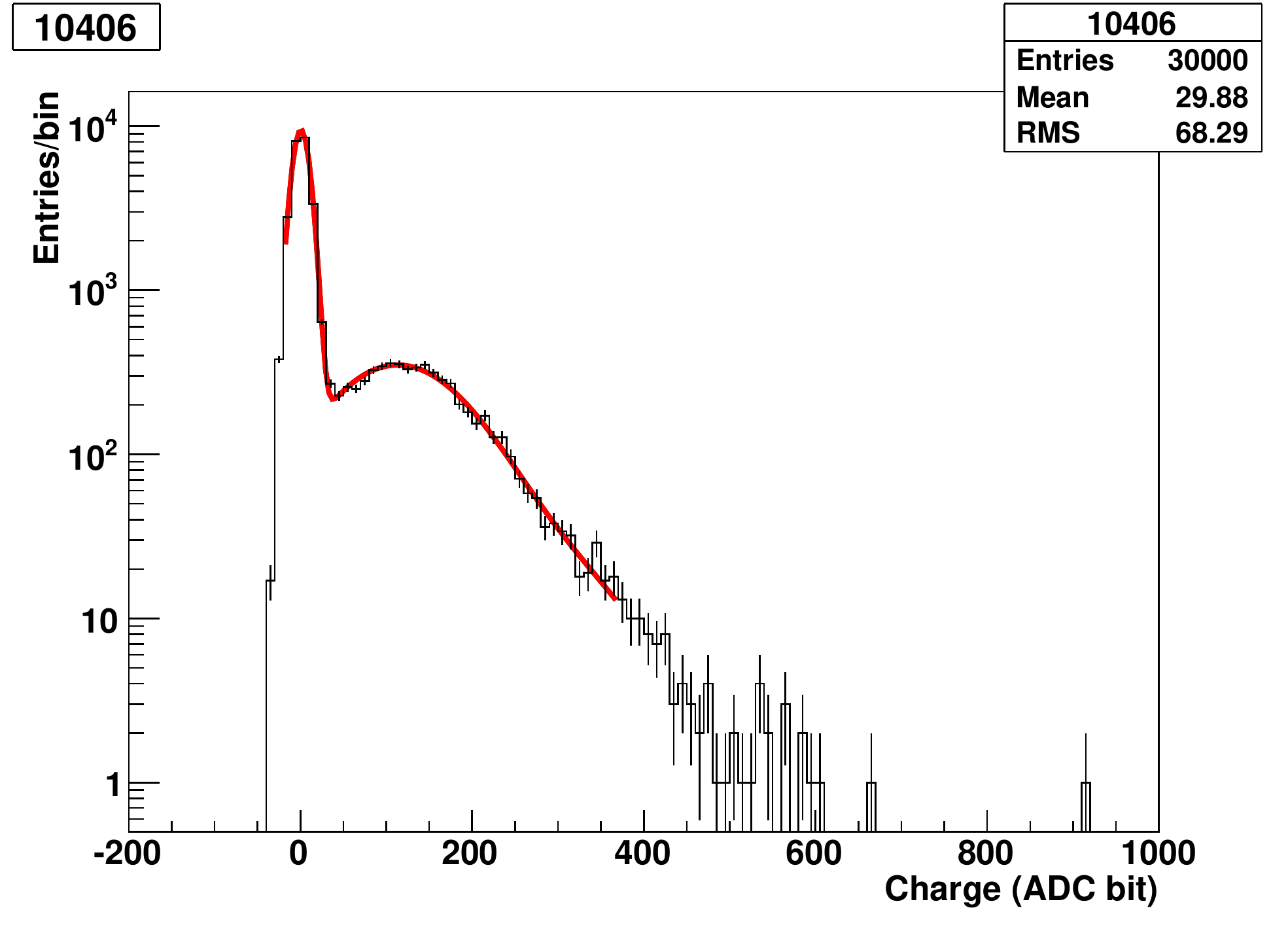}
\includegraphics[width=0.45\textwidth]{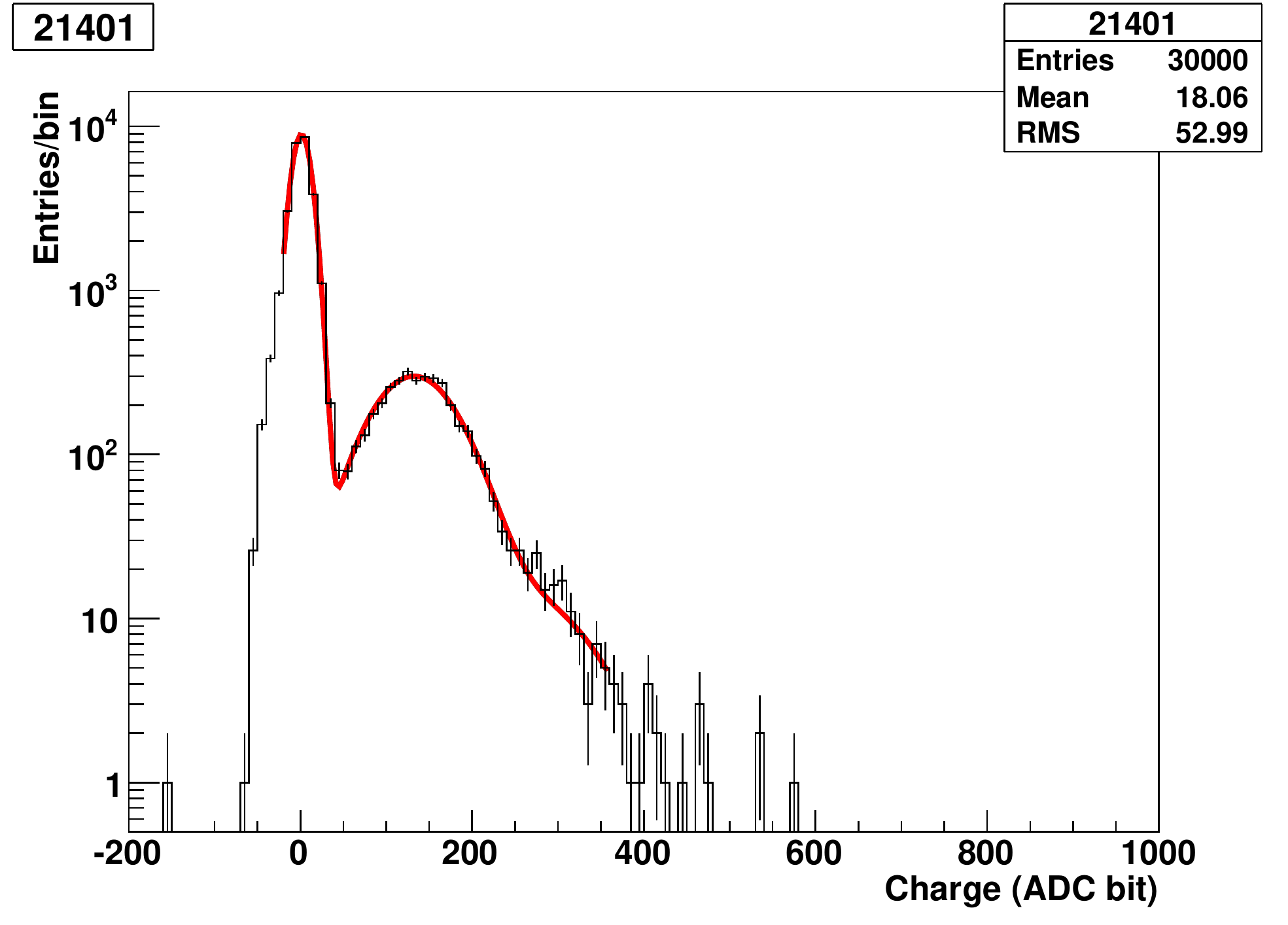}
\caption{Histograms of typical charge distributions for a R8520-406 (left)
and a R11410-MOD (right) PMT with fits overlaid (red curves).
The leftmost peak in each histogram is the pedestal.
The abscissa
are the integrated area of the SPE in ADC bits,
with a conversion of 1 bit = 2.75$\times10^{-3}$ pC before the
Phillips amplifier.}.
\label{fig:spe_spectrum}
\end{figure}

Low intensity LED runs were used to calibrate the gains of the PMTs.
The charge distribution for two typical PMTs are shown in
Fig.~\ref{fig:spe_spectrum}.
The gain and SPE resolution can be obtained by fitting the
SPE charge peak.
All PMT gains were adjusted to approximately $2\times10^6$ using an empirical
HV dependence of V$^\beta$, where $\beta$ is about 7(8) for the top (bottom)
bottom PMTs.
A typical resolution at this gain setting
for the R8520-406 PMTs is 60\% with the average at 71\%, biased by a few noisy
tubes. For the R11410-MOD PMTs, it is 39\%, including
the contribution from the phototubes as well as the entire electronics
chain.

\begin{figure}[!htbp]
\centering
  \includegraphics[width=0.45\textwidth]{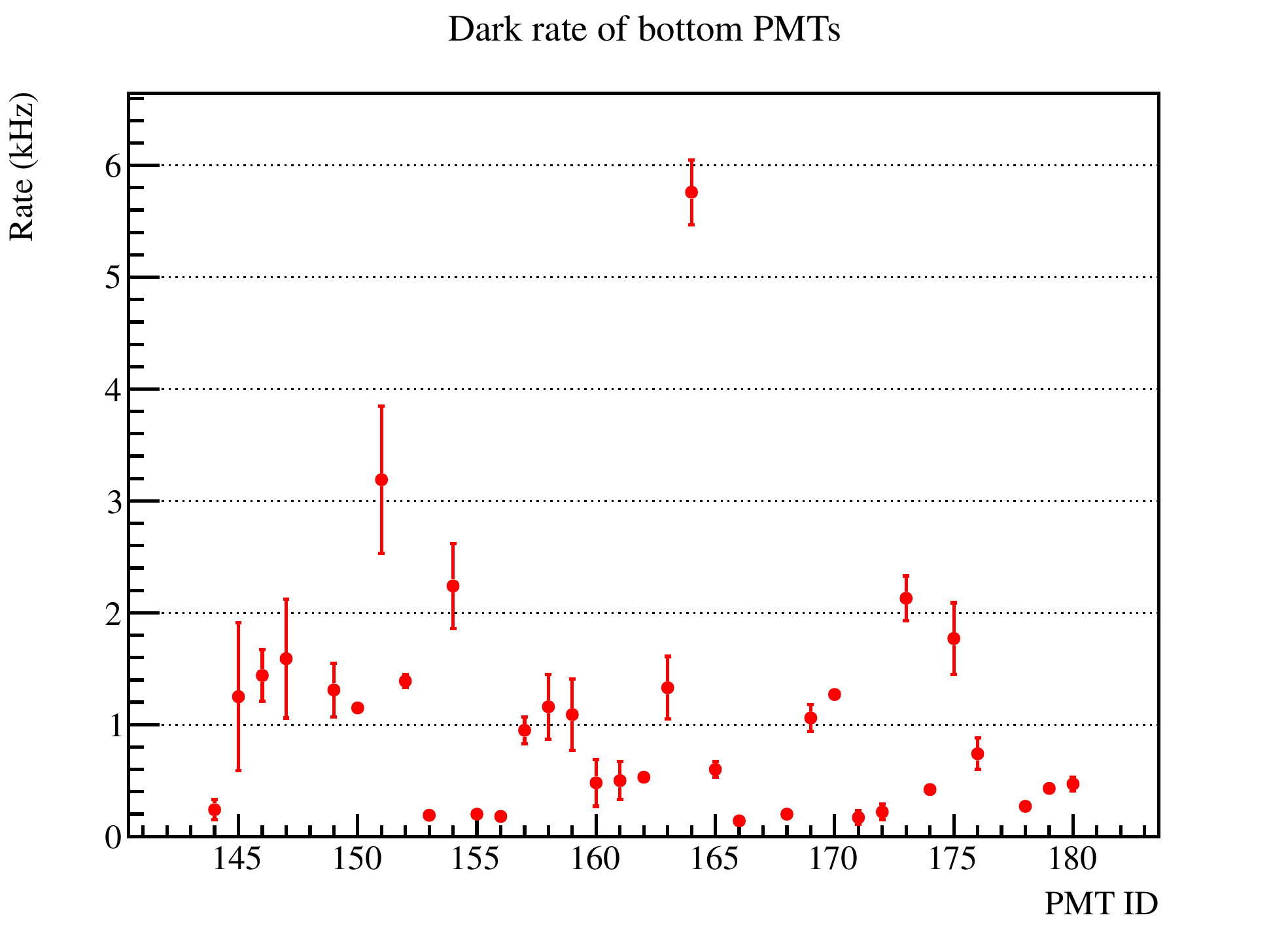}
  \includegraphics[width=0.45\textwidth]{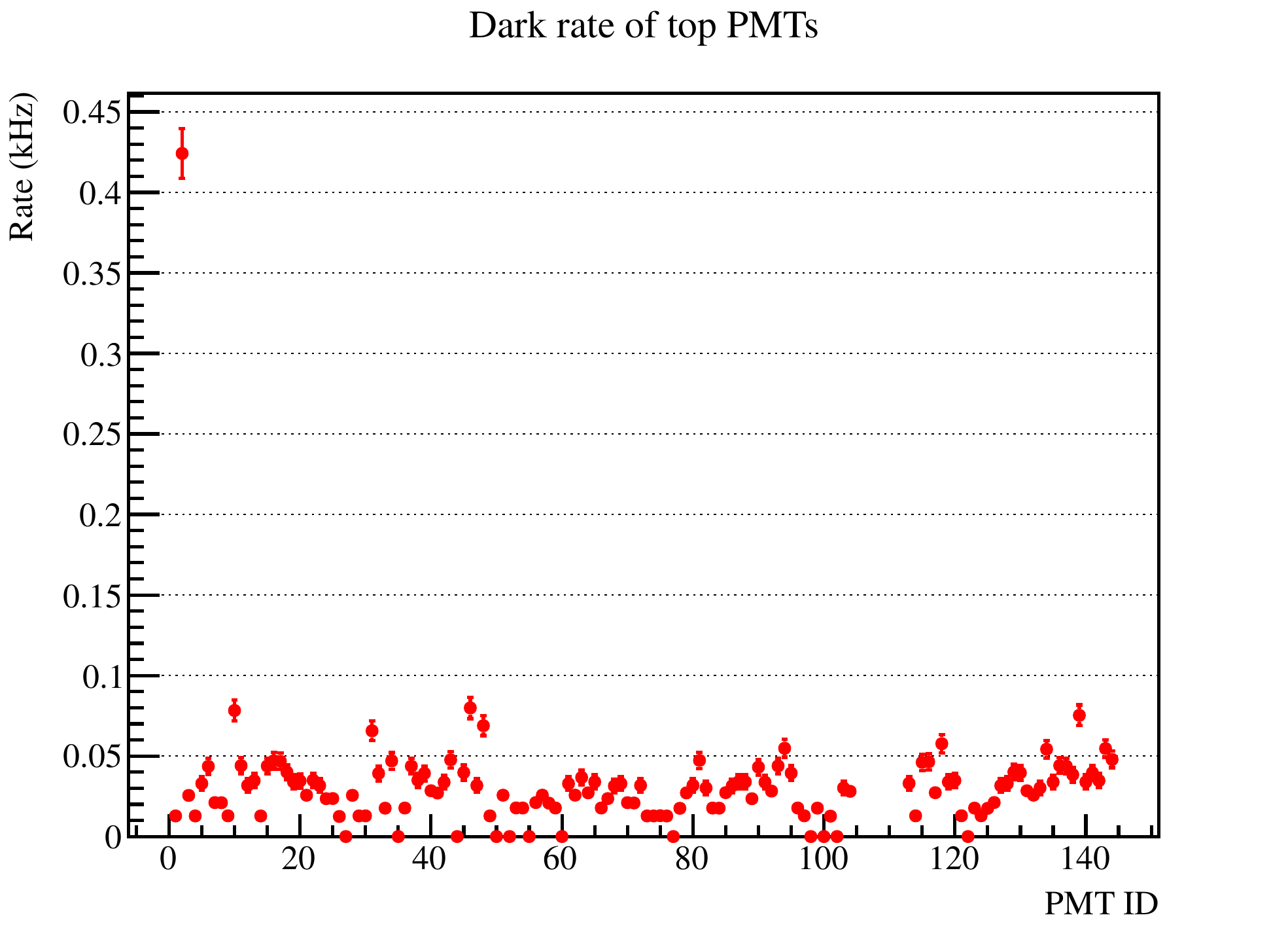}
\caption{Distribution of PMT dark rate in LXe for the bottom (left)
and top array. The uncertainties are computed based on the standard
deviations of the run-by-run fluctuations.}
\label{fig:dark_rate}
\end{figure}

The dark rate of PMTs in LXe are monitored periodically by taking DAQ runs
with random triggers. At a pulse finding threshold of 0.5\,photoelectron,
typical dark rate of a R8520-406 tube is 30\,Hz, and that for R11410-MOD
is 1\,kHz, with their distributions shown in Fig.~\ref{fig:dark_rate}.
Compared to the
initial data taken at room temperature, when the inner vessel
was filled with GXe, the top PMT dark rate dropped by a factor of 40,
while the bottom PMT dark rates only decreased by a factor of 2. To search for S1
pulses in the waveform, the offline software requires that at least
two phototubes fired within a window of approximately 100\,ns to form a signal.
Based on the random trigger run, the accidental probability due to PMT dark
rate within an S1 search window of 75\,$\mu$s is less
than 3\%.

The afterpulsing charge distribution and probability are also measured tube-by-tube. Low intensity LED runs were performed to produce SPEs in a narrow time window, and looked
for afterpulses in subsequent time windows. The average afterpulsing probability for the R11410-MOD tubes, computed as the ratio of hits between 0 to 4,000 ns (subtracted by the random hits estimated from the pretrigger hits) to the primaries, is 1.5\% in LXe and 1.7\% at room temperature. The same average probability for R8520-406 is 2.2\% (2.4\%) at room (LXe) temperature.


\section{Experimental Background}
\label{sec:background}

Controlling, understanding and reducing experimental background is crucial in dark matter direct detection experiments. As the target mass grows, this becomes ever more important. In this section, the various background sources to the PandaX experiment are considered, keeping in mind its evolution through the various stages to a ton-scale experiment. We briefly describe the Jinping underground lab, and the expected cosmic
ray and environmental radioactivity backgrounds. We further describe the passive PandaX shield, the  low-background counting station at CJPL, the special Kr purification tower, and results from Monte Carlo background simulations for the stage I experiment.

\subsection{CJPL and cosmic ray flux}

The Jinping deep underground lab, CJPL, emerged from a government-led project to construct two large hydropower plants next to and in the Jinping Mountain, Sichuan, China~\cite{jinping}, with a combined power output of 8.4\,GW. Jinping is located about 500\,km southwest of Chengdu, the capital of Sichuan province. It can be accessed either by car from Chengdu, or by a short flight to Xichang, followed by a 1.5\,hr car ride. CJPL was constructed jointly by Tsinghua University and the Yalong River Hydropower Development Company, Ltd. (EHDC) in Sichuan, China, in 2009.

\begin{figure}[!htbp]
\centering
  \includegraphics[height=2in]{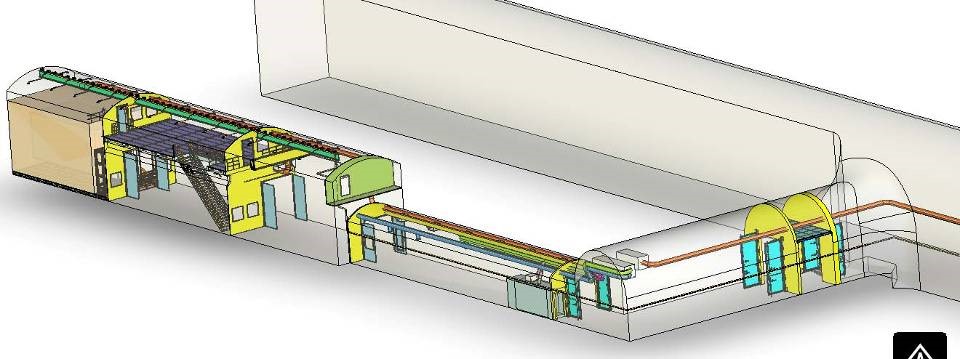}
\caption{Schematic picture of Jinping underground lab's L-shaped structure off traffic tunnel A.}
\label{fig:lab}
\end{figure}

The Jinping facility is located near the middle of traffic tunnel A, one of the two traffic tunnels that are maintained by the 21$^{\text st}$ Bureau of the China Railroad Construction Company. Its location, shown in Fig.~\ref{fig:lab}, is an ideal environment for low background experiments.
The facility is shielded by 2,400\,m of mainly marble, which is radioactively
quiet rock, as shown in
Table~\ref{tab:rock}. PandaX occupies the first 10\,m in a $6\times 6\times 40$\,m$^3$ cavern as depicted in
Fig.~\ref{fig:lab}. At a depth of 6,800 m.w.e. the muon flux was
measured~\cite{cite:CJPLcosmicray} at 62 events/(m$^2\cdot \text{year})$ (compared to about 100\,Hz/m$^2$
at sea level). Monte Carlo simulations show
that neutrons generated by those muons in the detector, the shielding
material, and the surrounding rock and concrete produce a nuclear recoil-like
background of less than 0.002/year in a 25\,kg detector. Natural radioactivity
from radon gas in the the lab is reduced to about 10\,Bq/m$^3$ by
flushing dry nitrogen gas through the passive shield, and contributes at negligible levels to the overall background.

\begin{table}[!htbp]
\centering
  \includegraphics[height=1.5in]{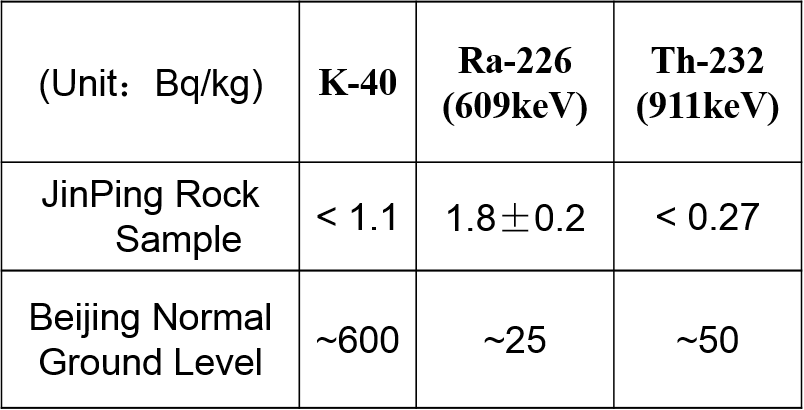}
\caption{Radioactivity of the rocks surrounding the CJPL lab.}
\label{tab:rock}
\end{table}

\subsection{Passive shield}

The PandaX passive shield was built to attenuate neutrons and gamma rays from environmental materials such as cavern wall rocks and concrete. It is shaped as an octagonal structure with a 316-cm width and a 368-cm height. It encapsulates a cylindrical shape of 124\,cm in
diameter and 175\,cm in height. The innermost layer of the shield is a cylindrical OFHC copper
vessel. Surrounding the shield structure is a steel platform that supports the cryogenic system and
the electronics. The total weight of the passive shield is 93\,tons, composed of 12\,ton copper, 58\,ton lead, 20\,ton polyethylene, and 3\,ton steel.

The shield was designed with two important requirements: (i) allow less than one neutron or gamma  induced background event per year in the energy region useful
for dark matter detection, and  (ii) satisfy the space constraint of CJPL, since the space allocation for PandaX was limited to a length of about 10\,m, and a width of 3.5\,m.

\begin{figure}[!htbp]
\centering
  \includegraphics[width=0.45\textwidth]{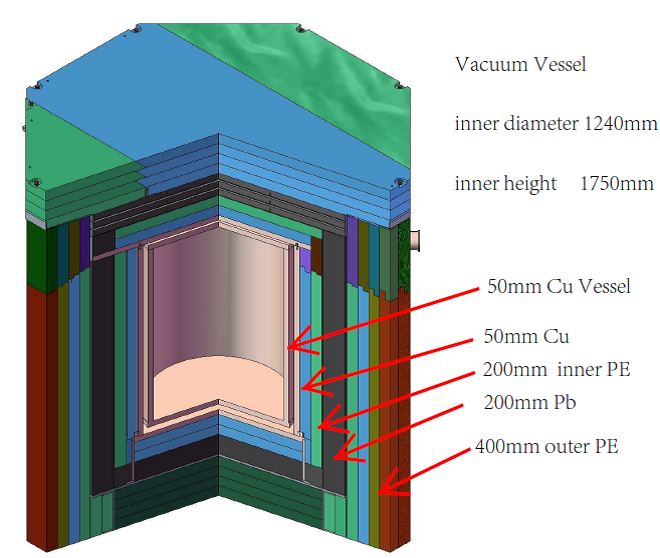}
   \includegraphics[width=0.45\textwidth]{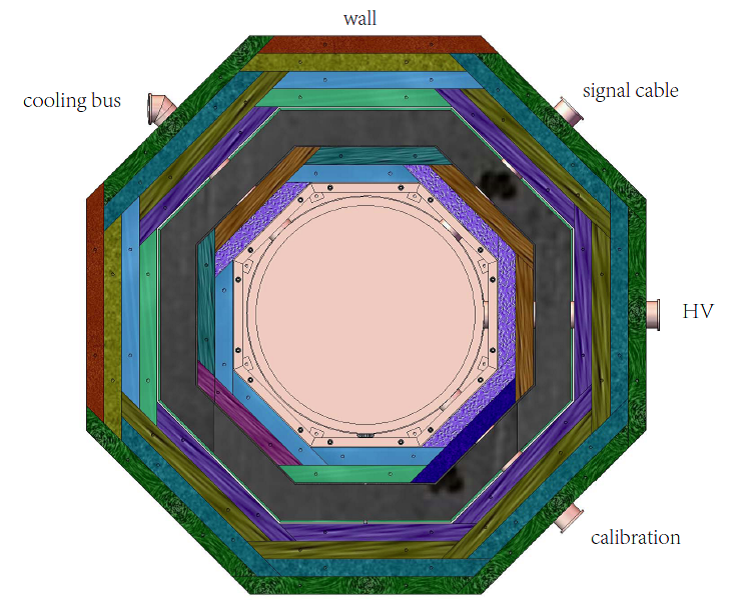}
\caption{Left panel: Passive shield cross sectional view. Right panel: Top view of the shield.}
\label{fig:sectionview}
\end{figure}

To satisfy these requirements, the shield was made of a 5 layered structure of, starting from the outside, 40\,cm polyethylene, 20\,cm lead,  20\,cm polyethylene, and 5\,cm copper. The innermost layer is a cylindrical 5-cm thick copper vessel, used both as the last layer of the shielding and as the wall of the cryostat.
The cross-sectional and top views of the shield are shown in Fig.~\ref{fig:sectionview}.
The horizontal leveling of the copper vessel is adjustable with two rotary degrees of freedom
with the range $±0.25^\circ$, used to adjust the leveling between liquid xenon surface
and TPC grids.
In the vertical direction, the maximum height of the crane's lifting hook is 510\,cm, which must accommodate
the length of the lifting rope (50\,cm), the height of the inner vessel plus the outer vessel, and the height of the shield base (~90\,cm).
Thus the maximum height of the outer vessel is about 185\,cm. In the horizontal direction,
after subtracting the shield thickness and construction space, the diameter of the outer vessel is limited to 135\,cm.
There are four 6-inch side openings around the outer vessel as shown in Fig.~\ref{fig:sectionview},
designed for the cooling bus tubes, the signal cables, and the high voltage and calibration feedthroughs.

Because polyethylene and lead have poor mechanical strengths, a steel structure had to be installed to support the weight of the shield. To contain radioactivity, most of the steel is used outside the lead shield.
Each side of the polyethylene plate is made with 45$^\circ$-sloped edges, and the small gaps between the adjacent plates are tangential to the inner space to prevent neutron leakage.



The shield covers are easily removal for detector installation and maintenance. They consist of one copper cover, one assembly cover, three lead covers and two outer polyethylene covers.
In the cover design, the payload of the crane is an important factor. The assembly cover weighes about 2,800\,kg, consisting of an outer top copper plate, an inner polyethylene layer, and some lead bricks. The lead cover is comprised of a steel box
and 14 layers of lead plates with 5\,mm thickness.  The steel boxes are mechanically rigid enough such that they can be lifted by the crane. Each steel box weighes about 3.5\,tons. The outer polyethylene cover is divided into two parts due to the space limitations in the lab.

\subsection{Counting station}
\label{Sec:counting-station}

The reach the required sensitivity for PandaX, stringent requirements are placed on the radioactivity levels of most detector components and materials. According to Monte Carlo simulations, the radioactive elements (we are mostly concerned with $^{238}$U, $^{232}$Th, $^{137}$Cs, $^{60}$Co, and $^{40}$K) in most central components should not exceed activity levels of mBq per piece or per kg. A counting station was constructed at CJPL to select materials that meets this requirement.

\begin{figure}[!htbp]
\centering
  \includegraphics[width=0.45\textwidth]{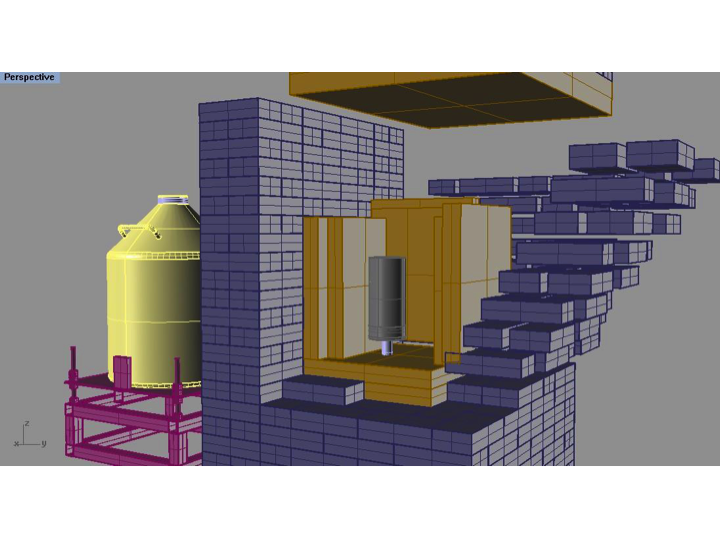}
  \includegraphics[width=0.45\textwidth]{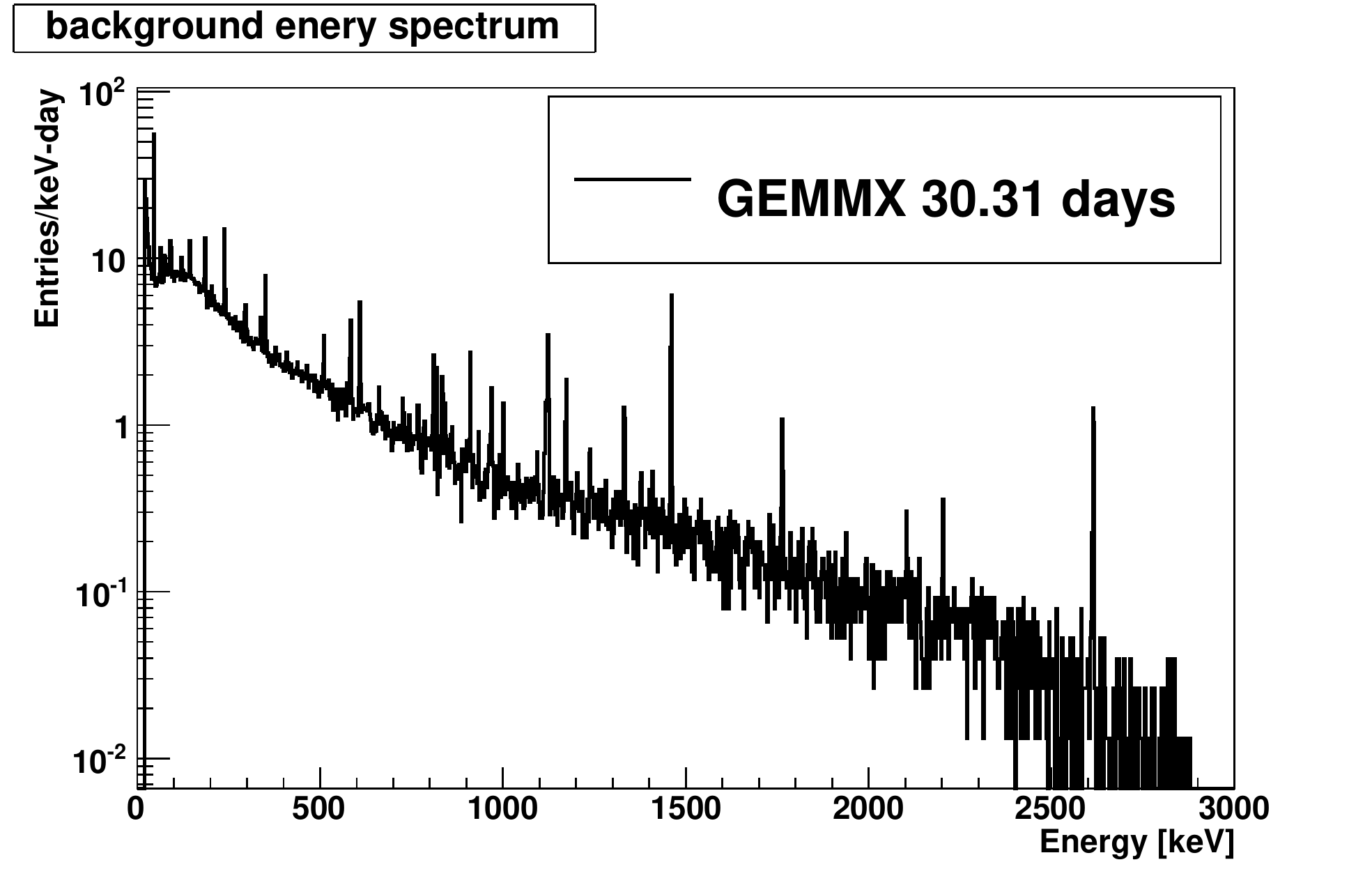}
\caption{Left panel: Illustration of the HPGe and the shielding.
The HPGe (grey) is surrounded by several OFHC copper plates (yellow) with a total thickness of 10cm.
The 20-cm thick lead shielding (blue) is built with lead bricks, each brick with a size of
5$\times$10$\times$20\,cm$^3$. Right panel: Background energy spectrum accumulated over a one-month period.}
\label{fig:counting1}
\end{figure}

An illustration of the counting station is shown in Fig.~\ref{fig:counting1}.
It houses an Ortec GEMMX low background, high purity germanium detector (HPGe), model 94100-HJ-LB-C-108. It has a relative efficiency of 175\%
and a resolution of 2.3\,keV (FWHM) at 1.33\,MeV. The signal is read out by an Ortec DSEPC502 DAQ system with a shaping time of 12\,$\mu$s.
The HPGe is shielded by 10-cm thick OFHC copper and 20-cm thick pure lead.
While the radioactive elements $^{226}$Ra($^{238}$U), $^{228}$Th($^{232}$Th), $^{137}$Cs and $^{60}$Co in the copper are all well below 1\,mBq/kg, $^{40}$K is relatively high, at 4$\pm$1\,mBq/kg, most probably due to sample surface contamination during counting.
The $^{210}$Pb level in the lead
ranges from 110 to 290\,Bq/kg. The counting chamber has a cross section of 20$\times$20\,cm$^2$ and is 34\,cm high (including the HPGe).
It is continuously flushed with dry nitrogen gas at 7--10\,L/min. Since the counting station is under operation at the CJPL underground laboratory, no anti-cosmic-ray system is needed.

The background energy spectrum, as shown in the right panel of Fig.~\ref{fig:counting1}, indicates traces of $^{210}$Pb, $^{228}$Th($^{232}$Th), $^{226}$Ra($^{238}$U)
and $^{40}$K.
If we assume that all $^{226}$Ra photon lines originate from radon gas in the counting chamber, the $^{222}$Rn level is 1\,Bq/m$^3$ and constitutes one of the dominant background sources.

Many detector components were counted, including the three-inch and the one-inch PMTs. Some of the results are listed in Table~\ref{table:counting_results}.
A very high level of $^{226}$Ra($^{238}$U) is found in the stainless steel sample from the inner vessel.
For PandaX stage-II, a new inner vessel will be built from more radiopure stainless steel.
\begin{table}[th!]
\center
\begin{tabular}{|c|c||c|c|c|c|c|c|}
\hline
samples  & unit    & $^{226}$Ra           & $^{228}$Th     & $^{40}$K         &$^{137}$Cs          & $^{60}$Co          & $^{235}$U    \\ \hline
PMT8520 & mBq/PMT  & $<$0.11         & $<$0.08   & 9.8$\pm$1.0 & 0.20$\pm$0.05 & 0.50$\pm$0.05 & $<$0.12 \\ \hline
Base 8520& mBq/base& .07$\pm$0.2     & $<$0.32   & $<$2.4      & 0.4$\pm$0.1   & $<$0.07       & $<$0.18 \\ \hline
PMT11410& mBq/PMT  & $<$0.72         & $<$0.83   & 15$\pm$8    & $<$0.31       & 3.4$\pm$0.4   & 1.4$\pm$1.2 \\ \hline
Base 11410&mBq/base& 1.1$\pm$0.1     & 0.16$\pm$0.15 & $<$1.3  & 0.33$\pm$0.08 & $<$0.06       & 0.4$\pm$0.2 \\ \hline
PTFE samples& mBq/kg & 2.3-33        & $<$1-$<$14 & $<$6 - 89  & 6 - 68        & $<$0.3 - $<$6 & $<$3 - $<$8 \\ \hline
SS sample & mBq/kg & 118$\pm$3       & 17$\pm$4  & 53$\pm$14   & $<$0.9        & 6.0$\pm$0.8   & $<$8     \\ \hline
\end{tabular}
\caption{
Radioactivity of the PMTs and some other detector components.
The PTFE samples are used for the TPC construction. The stainless steel (SS) sample
is from the stage-I inner vessel.
}
\label{table:counting_results}
\end{table}

\subsection{Xenon purification system}

Commercially supplied xenon contains a Kr contamination of about 10\,ppb. Therefore, a tower to remove
Kr from xenon using standard distillation methods was designed and constructed at SJTU.
The distillation tower was designed to fulfill the following requirements:
The amount of Kr in the purified xenon should be reduced by up to three orders of magnitude compared to that in the original xenon. The collection efficiency of xenon should be 99\%. The processing speed of the system should be 5\;kg xenon per hour, so that 840\,kg xenon gas can be purified in a week.
Xenon should be fed into the distillation system as a gas.

The McCabe-Thiele (M-T) method was used to design the distillation column~\cite{mt}.
The original xenon gas flows into the distillation column through a feeding point, and the purified xenon is exhausted from the re-boiler. The concentration of Kr in purified xenon is lower than that in original xenon. The off gas is exhausted from the condenser which contains the highest concentration of Kr. The reflux of xenon is controlled both by the heating power provided by the re-boiler, and the cooling power provided by the GM cryo-cooler.

\begin{figure}[!htbp]
\centering
  \includegraphics[width=0.5\textwidth]{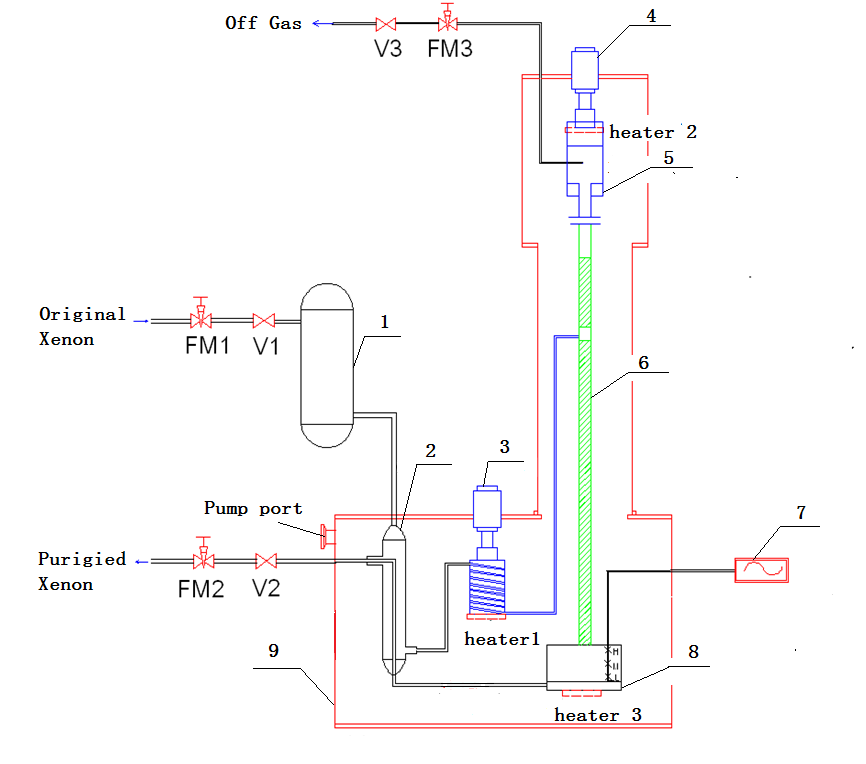}
  \includegraphics[width=0.35\textwidth]{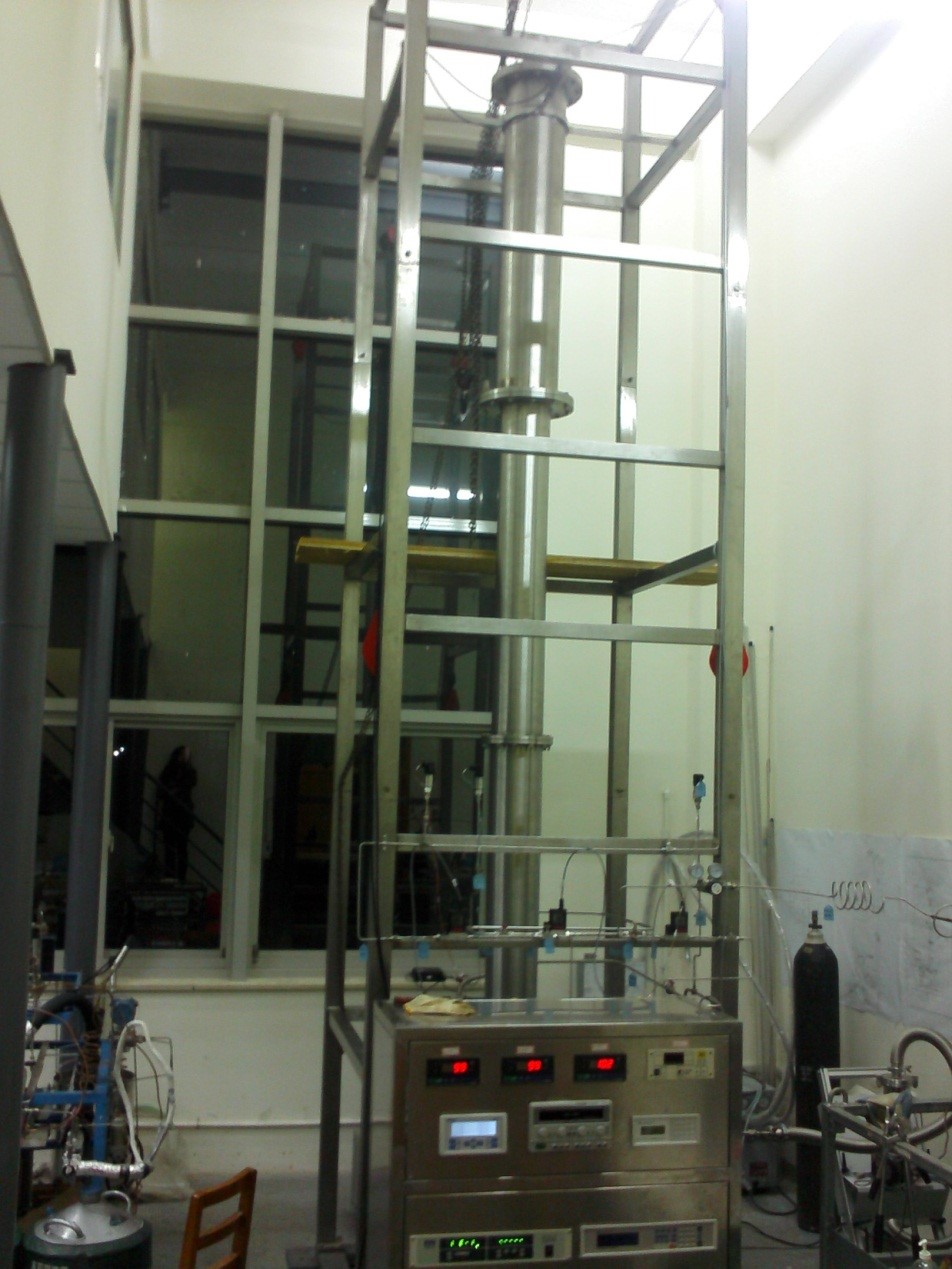}
\caption{Left panel: Schematic overview of the technological process of the cryogenic distillation system, 1--Getter, 2--Heat exchanger, 3--Cryo-cooler 1,
4--Cryo-cooler 2,   5--Condenser,   6--Distillation column,   7--Liquid meter,   8--Re-boiler,  9--Vacuum jacket. Right panel: Photograph of the actual
distillation system.}
\label{fig:techpro}
\end{figure}

The flow diagram of the distillation system is illustrated in Fig.~\ref{fig:techpro}. The original xenon gas pressure is set to 215\,kPa before entering the distillation system. The flow rate of the original xenon gas is controlled by both the needle valve FM1 and the flow controller. The original xenon gas flows through the getter, and through the heat exchanger, which is used to pre-cool the xenon gas. For this purpose, the purified xenon gas extracted from the distillation apparatus is used as the cooling medium~\cite{mk}. After the pre-cooling process, the temperature of the xenon gas is reduced to 192\,K sequentially by GM cryo-cooler 1, and the cooled xenon gas is fed into the distillation column. After being fed in, the xenon gas is further cooled down by the liquid xenon that drops down from the condenser to reach gas-liquid equilibrium at the packings. The xenon gas in the condenser is liquefied by GM cryo-cooler 2. The output cooling power is controlled by electrical heaters installed at cryo-cooler 2 and by the PID temperature controller. The temperature is kept at 179.5\,K, which is controlled by silicon diode temperature sensors. The saturated xenon gas with the highest concentration of krypton, named the ``off gas'', is collected from the condenser, while the flow rate is controlled by the needle valve FM3 and the flow controller. The re-boiler is a cylindrical stainless steel vessel with a height of 318\,mm and a diameter of 273\,mm, respectively. Some of the liquid xenon in the re-boiler is vaporized and returns to the distillation column for recycling, and the other liquid xenon, called ``purified xenon'', is collected in a collection bottle~\cite{slutsky} via the heat exchanger. The flow rate of the purified xenon gas is controlled by the needle valve FM2 and by the flow controller. The purified xenon and off gas are collected in stainless steel bottles cooled by liquid nitrogen.

A photograph of the distillation column is shown in the right panel of Fig.~\ref{fig:techpro}.
The required purity level of the purified xenon determines the height of the distillation column. In general, the higher the distillation column, the higher the purity. To guarantee the required purity level, the total height of the distillation column is set to 4\,m. The lengths of the rectifying and the stripping sections of the distillation column are 1.9\,m and 2.1\,m, respectively. The inner diameter of the distillation column is 0.08\,m. The cryogenic distillation column is thermally insulated by high vacuum multilayer insulation, and the vacuum is maintained at about $6\times 10^{-3}$\,Pa by a vacuum pump. The total heat leakage of the column is less than 6\,W.
The condenser at the top of the distillation tower is designed as funnel-type.
Its output power is adjusted by three heaters installed at the cold head of the cryo-cooler, and its temperature is regulated at 179.5\,K.
The electrical heater is adopted in the re-boiler at the bottom of the distillation tower, and the temperature of the re-boiler can be controlled at 180\,K. A capacitive liquid meter is installed in the re-boiler to monitor the amount of liquid xenon.

Stainless steel 304 was used as the material of the distillation tower.
A new type of high efficient structured packing PACK-13C~\cite{li} was adopted for this distillation column, which reflects the advantages of structured packing and random packing with the specific surface area of 1,135\,m$^2$/m$^3$. The PACK-13C structure was fabricated by suppressing double-layer silk screen, made of stainless steel, into a corrugated shape.

This distillation system has been used to purify 500\,kg xenon gas that contained ppb levels of  Kr concentrations. The concentration of Kr in the off gas was  about 10$^{-6}$\,mol/mol, which is consistent with the transfer of the majority of the Kr in the original xenon sample to the off-gas. Assuming mass conservation, the concentration of Kr in the purified xenon is expected to be at the ppt level, which satisfies the requirements for G2 dark matter experiments.

\subsection{Background simulation}

The three main background sources for the PandaX experiment are (i) the detector components, (ii) the CJPL lab concrete walls (including the rock), and (iii) the cosmic ray muons (including induced secondary particles). To estimate the background levels,
a Geant4 based Monte Carlo simulation package was constructed with the lab and detector geometry implemented.
To simulate the energy deposited inside the sensitive liquid xenon volume from various background sources, three selection cuts were applied to the simulated events.
First, a fiducial volume (FV) was defined in the center of the liquid xenon sensitive volume, with a radius
of 25\,cm and a height of 5\,cm. The fiducial volume was surrounded in all directions by at least a 5-cm thick layer of additional liquid xenon.
With the so-called FV cut, events that deposit more than 0.1\,keV energy outside the FV are rejected.
Second, under the assumption that the TPC position reconstruction resolution is 3\,mm,
only single-site events are accepted. Those are events with no energy deposits above 1\,keV and separation of more than 3\,mm.
Third, for the so-called electron recoil events, i.e., events with energy deposit from gamma or electron, but not from neutron interactions, an additional 99.5\% rejection efficiency from the S2-S1 selection is assumed.

The number of background events per year, within a 5 to 15\,keV window is summarized in Table~\ref{table:simulation_results}. Other copper components include the shaping rings, the copper plates to hold the top and bottom PMT arrays, and the copper filler.
\begin{table}[th!]
\center
\begin{tabular}{|c|c|}
\hline
background source  & \# bkgd. events / year \\ \hline\hline
cosmic muon        & $<$0.0015            \\ \hline
CJPL concrete wall &  $<$0.009     \\ \hline
PMTs and bases     & 4.4           \\ \hline
PTFE reflectors    & 0.44 (gamma) + 0.66 (neutron) \\ \hline
inner SS vessel    & 3.2 (gamma) + 0.24 (neutron) \\ \hline
Cu components      & 1.2      \\ \hline
outer Cu vessel    & 0.39       \\ \hline
Rn (1 Bq/m$^3$)    & 0.007     \\ \hline
\end{tabular}
\caption{
Number of background events per year from various detector components and sources. The contributions from PMTs etc are for the stage-I experiment.
}
\label{table:simulation_results}
\end{table}

At present,  CJPL  has the largest overburden and the correspondingly
lowest cosmic muon flux at $(2.0\pm0.4) \times 10^{-10}$/(cm$^2\cdot$s)~\cite{cite:CJPLcosmicray}.
Ten thousand muons were simulated and no single event survived the three selection cuts.
To increase simulation statistics, 1 million muon-induced neutrons following the energy distribution parameterized
in Ref.~\cite{cite:Mei_Hime_underground_bkg} were simulated and only 1 event survived the FV and single-site cuts.
After all selection cuts, the cosmic muon induced background within 5 to 15\,keV is less than 0.0015 per year.

The CJPL concrete walls contain (1.63$\pm$0.17)\,kBq/kg of $^{226}$Ra,
(6.5$\pm$0.9)\,Bq/kg of $^{232}$Th and (19.9$\pm$3.4)\,Bq/kg of $^{40}$\,K~\cite{cite:CDEXintroduction}.
The gamma induced background events within the 5 to 15\,keV window is less than 0.001 per year, thus negligible for PandaX stage-I.
The neutron flux from the CJPL concrete walls from  ($\alpha$,n) and  spontaneous fission of U and Th was simulated together with the SOURCES-4A~\cite{cite:source4a} package and Geant4,
and is $2.0\times 10^{-6}$/(cm$^2$$\cdot$s).
The induced background within the 5 to 15\,keV window is less than 0.008 per year.

The simulations have shown that the dominant background originates from the detector components. The gamma-induced background event rate is 18.1\,mDRU after the single-site and fiducial volume selections, including 8.5\,mDRU from the PMTs and 6.1\,mDRU from the stainless steel inner vessel.
If we assume the Kr in the liquid xenon is at the 70\,ppt level, it will contribute an additionally 2.8\,mDRU.


\section{Electronics and Data Acquisition System}

The electronics and data acquisition system takes the signals detected by the
PMTs 
and, after a sequence of signal processing,
produces and records raw data for physics analysis. The main
functionality of the system includes signal amplification, waveform
sampling and digitization, trigger and event readout, as well
as system monitoring.

The PandaX detector is
instrumented with 180 PMTs to detect
scintillation photons.  A typical physics event 
consists of
two time-correlated signals, S1 and S2. Combining the signals recorded by all photomultiplier tubes, the
typical range of the S1 signal for low energy nuclear
recoil events extends from a few to a few tens of
photoelectrons. The charge of the corresponding S2 signal is typically
one hundred times larger, 
with a width of a few $\mu$s. For stage I, the maximum electron drift length
is 15\,cm. Assuming a drift speed of 2\,mm/$\mu$s, the
maximum separation between S1 and S2 is 75\,$\mu$s.

A channel-by-channel digitizer system (flash ADC or FADC) was chosen for PandaX. The electronics and data acquisition system have to satisfy the following requirements: (i) the digitizers have to be able to measure the charge of the SPEs
accurately. For a gain of 2$\times10^{6}$, the raw SPE signal has an amplitude of about 2\,mV and a width of 10--20\,ns. This sets the specification of the digitizer to have a sampling rate of
at least 100\,MS/s and to have a charge resolution of 14\,bit. Pre-amplification
is also desired to boost the raw signal size; (ii) the trigger has to be highly efficient for low energy nuclear recoils. For these events, since both the S1 and S2 amplitudes are small, a charge
trigger has to be used with a threshold lower than 200\,PE (on S2); (iii) each readout window needs to be larger than 150\,$\mu$s so that the S1-S2 event pair is captured in a single readout regardless of which
generated the trigger; and  (iv) the event rate, together with the event size, determines the bandwidth
for data transfer. Based on simulations, an event rate
with a 1\,keV$_{ee}$ threshold is about 3\,Hz. It is therefore required that the
DAQ is capable of taking more than 10\,Hz continuously without deadtime.

\subsection{Electronics design}
\begin{figure}[!htbp]
   \centering
   \includegraphics[width=0.95\textwidth]{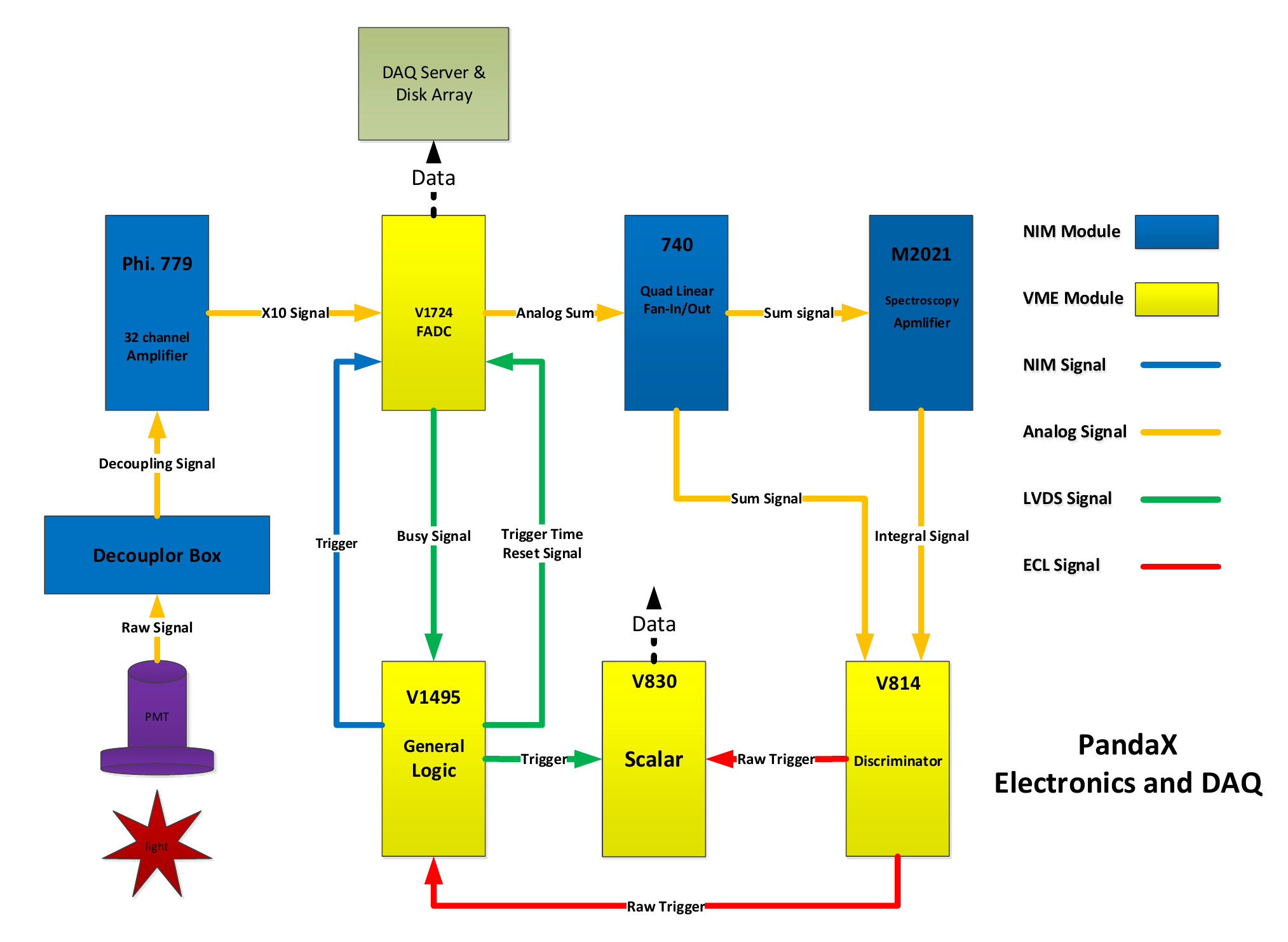}
   \caption{Schematic layout of the electronics and DAQ system.}
   \label{fig:schematic_EnDAQ}
\end{figure}

The design of the PandaX electronics is very similar to that of previous expriments~\cite{XENON100, LUX}.
The hardware components were commercially available, while the
DAQ software was developed in-house. The schematic layout of the electronics system is shown in
Fig.~\ref{fig:schematic_EnDAQ}.

The PandaX eletronics system is a mixed VME and NIM system.
The NIM amplifier modules (Phillips 779) take PMT signals from the decoupler, amplify them by a factor of 10, and feed them to FADC modules. The CAEN V1724 FADCs perform waveform sampling of the input signals (100\,MS/s, 14\,bit) and provide digitally summed signals at a 40\,MS/s frequency (Esum). When receiving trigger signals, the analog samples are recorded in the event buffer and ready to be read out by the DAQ computer via optical link.  
The signals of the digital sum from all FADC modules are summed via FAN-IN/OUT modules to form the Esum signal, which is integrated by a spectroscopy amplifier (ORTEC 575A) with a 1.5\,$\mu$s shaping time, producing a signal for which the height scales with the total charge of all channels. This charge signal is fed into a CAEN V814 discriminator. Tests have shown that the threshold can be lowered to about 150-200\,PE, sufficient to trigger S2 for low energy nuclear recoil events. Efforts are continuing to further reduce that threshold. The resulting \emph{Trig} signal is fed into a general logic module (CAEN V1495), which makes a trigger decision together with all \emph{Busy} signals from the FADCs as the VETO. The resulting output signal is the main trigger for the FADC readout window. The scaler module (CAEN V830) is used to monitor the system performance, as for example the trigger efficiency and the live time. 
Fiber optical links connect the VME electronics to the DAQ computer.
When there are data ready in the FADCs' buffers, the DAQ program will read out the data from all FADCs units via the so-alled MBLT cycle. The data segments are built as physics events and are written  to a disk array attached to the computer for permanent storage.

\begin{figure}[htbp]
  \centering
   \includegraphics[width=0.8\textwidth]{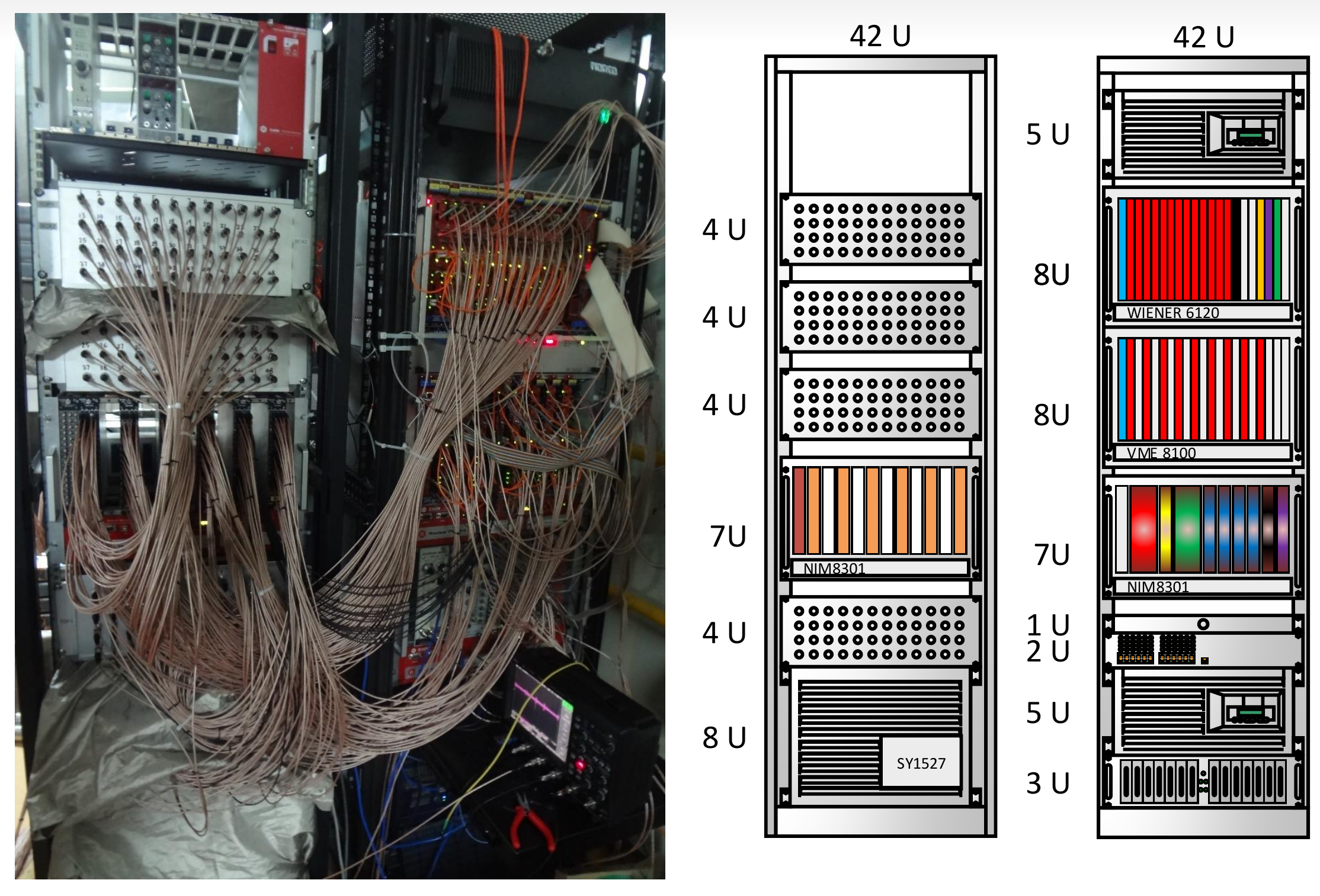}
   \caption{Photograph of the electronics system in Jinping (left) and a schematic layout of the rack components (right). The decoupler units (4x4U) as well as the HV supply crate can be found in the left rack, together with NIM crate hosting the Phillips pre-amplifiers. The right rack contains the VME electronics and the trigger logic.}
   \label{fig:rack}
\end{figure}

A photograph of the electronics system at CJPL and a schematic layout of the rack components are shown in Fig.~\ref{fig:rack}.

\subsection{Data acquisition and online data processing software}
The DAQ software was written in C++ using the software library
provided by CAEN. The basic logic flow uses a simple framework which allows to
configure the DAQ via an input XML file and
start/end a run. The data is transferred from the FADC to the
DAQ computer via optical bridge using the MBLT protocol to minimize
the system deadtime. The raw data file is in custom binary format,
encoding all waveforms. Waveform samples consistent with the baseline
are suppressed with a threshold of 20 ADC bits using the FPGA algorithm
provided by CAEN (so-called baseline suppression). This suppression leads to about a factor 8 reduction in
data size.

\begin{figure}[!htbp]
\centering
\includegraphics[width=0.65\textwidth]{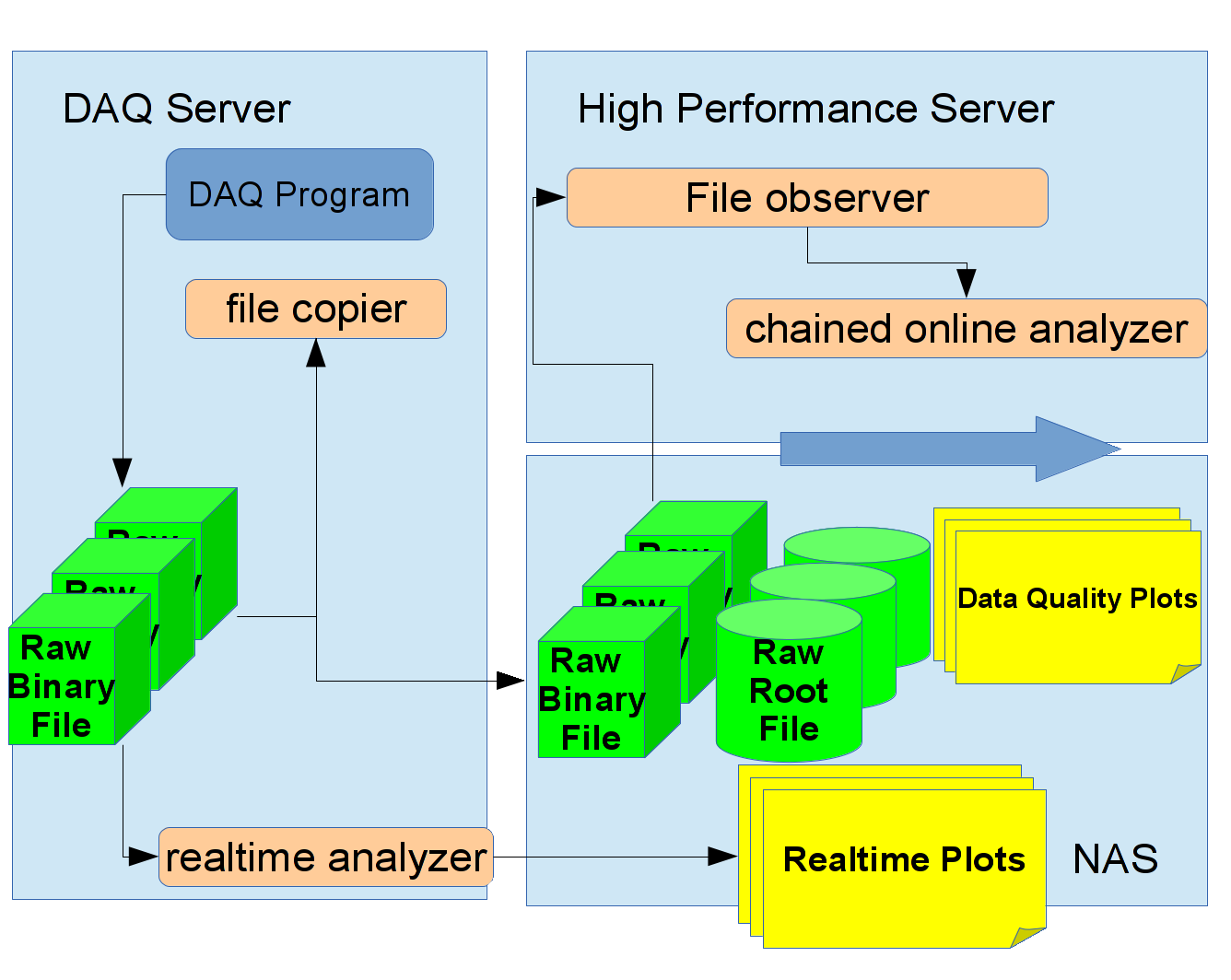}
\caption{The architecture of the data acquisition and online monitoring
system.}
\label{fig:daq3}
\end{figure}

The downstream data processing and monitoring scheme is illustrated in
Fig.~\ref{fig:daq3}.
To achieve real time monitoring of the system performance,
a separate process parses the output file on the fly and produces
data quality figures.
%
The DAQ computer is located inside the underground laboratory, with binary
data written to a local disk array during data taking. A real-time analyzer program reads the data on the fly and generate
plots of hit map, wave forms and trigger rate on a web page.
A file copier program, running on the same
server, copies the new data to a remote file system provided by a net
attached storage (NAS) server outside the laboratory, connected to the
underground lab via fiber links. The NAS server
is attached to a multi-core computing server. A file tracker program will
invoke an analysis chain once raw binary file copying is completed
to start converting the raw binary file into a root format, analyzing
the data and generating more detailed data quality plots, all viewable via
a web browser.

The data transfer rate was tested throughout the entire processing chain
to an upper limit of 60\,MB/s, corresponding roughly to a trigger rate of
100\,Hz with a 200\,$\mu$s readout windows on all channels, including baseline suppression
enabled.

\section{Summary}

In this paper, we report the design and performance for a dark matter direct detection search experiment operating in xenon dual-phase mode in the Jinping underground laboratory in China. The cryogenic system, the gas circulation system, the gas purification system, the level adjusting mechanism, the TPC high voltage, the PMT and the S1 and S2 signals have all been tested in two engineering runs, and performed as expected.  The experiment entered science data collection in late March 2014, and is expected to report results later this year. After completion of stage I,   the detector will be upgraded to accommodate a larger target mass.

This project has been supported by a 985 grant from Shanghai Jiao Tong University, a 973 grant from
Ministry of Science and Technology of China (No. 2010CB833005),  a grant from National Science Foundation of China (No.11055003), and
a grant  from the Office of Science and Technology in Shanghai
Municipal Government (No. 11DZ2260700). The project has also been sponsored by Shandong University, Peking University, the University of Maryland,
and the University of Michigan.
We would like to thank many people including E. Aprile, X. F. Chen, C. Hall, T. D. Lee, Z. Q. Lin, C. Liu, L. L\"u, Y. H. Peng, L. Teng, W. L. Tong, H. G. Wang,
J. White, Y. L. Wu, Q. H. Ye, and Q. Yue for helps and discussions at various level. We are particularly indebted to Jie Zhang for his strong support and crucial help
during many stages of this project. Finally, we thank the following organizations and personnel for indispensable logistics and other supports:
the CJPL administration including directors J. P. Chen and K. J. Kang and manager J. M. Li, Yalong River Hydropower
Development Company, Ltd. including the chairman of the board H. S. Wang, and manager X. T. Chen and his Jinping tunnel management
team from the 21st Bureau of the China Railway Construction Co.

\end{document}